\documentclass[10pt]{article}
\pdfoutput=1

\usepackage{mathrsfs}
\usepackage{booktabs,tabularx,array,longtable}
\usepackage{setspace}
\usepackage[letterpaper,top=25mm,left=28mm,bottom=27mm,right=28mm]{geometry}
\usepackage{titlesec}
\titleformat*{\section}{\Large\bfseries}
\titleformat*{\subsection}{\large\bfseries}
\titleformat*{\subsubsection}{\bfseries}
\titleformat*{\paragraph}{\bfseries}
\usepackage{booktabs}
\usepackage{tikz}

\definecolor{survey-1}{HTML}{3366A4}
\definecolor{survey-3}{HTML}{237A45}
\definecolor{survey-4}{HTML}{B5A31B}
\definecolor{survey-5}{HTML}{DB566D}
\definecolor{survey-6}{HTML}{9A3576}
\definecolor{survey-7}{HTML}{54368A}
\definecolor{survey-8}{HTML}{B66A22}
\definecolor{survey-9}{HTML}{8D253B}
\definecolor{survey-11}{HTML}{CB4B28}
\definecolor{survey-12}{HTML}{717E2D}
\definecolor{survey-13}{HTML}{BE589E}
\definecolor{survey-14}{HTML}{227C8D}
\definecolor{survey-15}{HTML}{5046A5}
\definecolor{survey-16}{HTML}{B18A6A}
\definecolor{survey-18}{HTML}{72933C}
\definecolor{survey-19}{HTML}{66504B}
\definecolor{survey-20}{HTML}{DA8E86}
\definecolor{survey-22}{HTML}{6C87C6}

\newcommand{\SurveyDot}[1]{%
  \tikz[baseline=-.5ex]{%
    \fill[survey-#1] (0,0) circle[radius=2.9pt];
    \ifnum#1=20
      \fill[white] (0,0) circle[radius=1.25pt];
    \fi
    \ifnum#1=22
      \fill[white] (0,0) circle[radius=1.25pt];
    \fi
    \draw[black!45,line width=.18pt]
      (0,0) circle[radius=2.9pt];
  }%
}



\usepackage{soul}

\usepackage{mathtools}
 \usepackage{tcolorbox}
    \tcbuselibrary{listings,theorems}

\usepackage{upgreek}

\usepackage{empheq}
\usepackage{microtype}
\usepackage{relsize}
\usepackage[dvipsnames]{xcolor}
\usepackage{cite}
\usepackage{enumitem}
\usepackage{ifpdf}
\usepackage{lmodern}
\usepackage[T1]{fontenc}
\usepackage[ansinew]{inputenc}
\usepackage{amsmath}
\usepackage{amsthm}
\usepackage{relsize}
\usepackage{amssymb}
\usepackage{comment}
\usepackage{tikz}
\usepackage{graphicx}
\usepackage{color}
\definecolor{darkblue}{cmyk}{0.9,0.9,0,0}
\definecolor{darkgreen}{rgb}{0,0.55,0}
\usepackage[colorlinks=true,linkcolor=darkblue,citecolor=darkblue,urlcolor=darkblue]{hyperref}
\usepackage{epsfig}
\usepackage{epstopdf}
\usepackage{graphicx}

\makeatletter
\long\def\@makecaption#1#2{
  \vskip\abovecaptionskip
  \sbox\@tempboxa{{\captionfonts #1: #2}}
  \ifdim \wd\@tempboxa >\hsize
    {\captionfonts #1: #2\par}
  \else
    \hbox to\hsize{\hfil\box\@tempboxa\hfil}
  \fi
  \vskip\belowcaptionskip}
\makeatother

\def\las{\lambda_{\sf a}}

\newcommand{\barAbulk}{\bar A_{\mathsf{bulk}}}
\newcommand{\dAbulk}{\delta A_{\mathsf{bulk}}}

\newcommand{\Qdelta}{Q_\delta}
\newcommand{\Qbar}{Q_{\overline{\bulk}\,}}

\def\edge{{\sf edge}}
\def\bulk{{\sf bulk}}

\def\sfb{{\sf b}}
\def\lb{\l_\sfb}
\def\gb{\g_\sfb}

\def\EE{\mathbb{E}}
\def\PP{\mathbb{P}}
\def\bitem{\begin{itemize}}
\def\eitem{\end{itemize}}
\def\benum{\begin{enumerate}}
\def\eenum{\end{enumerate}}
\def\edge{{\sf edge}}
\def\bulk{{\sf bulk}}

\def\sfb{{\sf b}}

\definecolor{greencirc}{rgb}{0, 0.73, 0.5}
\def\thalf{{\tfrac12}}

\def\enumcom{\begin{enumerate}[label*=  \sf (\roman*),wide, labelwidth=!, labelindent=0pt]}

\def\ZZ{\mathbb{Z}}

\def\RR{\mathbb{R}}

\def\lz{L_{\cZ}}

\def\Lamz{\L_{\cZ}}

\def\vareps{\varepsilon}
\def\rhob{\bar\rho}

\def\sfa{\mathsf{a}}
\def\sfb{\mathsf{b}}

\def\cF{\mathcal{F}}

\def\L{\Lambda}
\def\half{{1\o2}}

\def\R{\mathbb{R}}

\def\x{\times}
\def\cB{\mathcal{B}}

\def\eps{\epsilon}

\def\hb{\overline h}
\def\Z{\mathbb{Z}}

\def\1{{\rm 1-loop}}

\def\cZ{\mathcal{Z}}

\def\c{\cite}

\def\cA{\mathcal{A}}

\def\cN{\mathcal{N}}
\def\cA{\mathcal{A}}

\def\c{\cite}

\def\vs{\vskip .1 in}

\def\G{\Gamma}

\def\o{\over}
\def\g{\gamma}
\def\D{\Delta}
\def\rar{\rightarrow}

\def\eqr{\eqref}

\def\i{\infty}

\def\ssec{\subsection}
\def\sssec{\subsubsection}
\def\sec{\section}

\def\foot{\footnote}
\newcommand{\es}[2] {\begin{equation} \label{#1} \begin{split} #2 \end{split} \end{equation}}
\newcommand{\e}[2] {\begin{equation} \label{#1} #2 \end{equation}}

\newcommand{\beq}{\begin{equation}}
\newcommand{\eeq}{\end{equation}}
\newcommand{\beqy} {\begin{eqnarray}}
\newcommand{\eeqy} {\end{eqnarray}}
\newcommand{\bsmat}{\begin{smallmatrix}}
\newcommand{\esmat}{\end{smallmatrix}}
\newcommand{\bmat}{\begin{matrix}}
\newcommand{\emat}{\end{matrix}}

\def\({\left(}
\def\){\right)}
\def\[{\left[}
\def\]{\right]}

\def\<{\langle}
\def\>{\rangle}

\def\a{\alpha}
\def\b{\beta}
\def\g{\gamma}
\def\G{\Gamma}
\def\d{\delta}
\def\D{\Delta}
\def\z{\zeta}

\def\k{\kappa}
\def\l{\lambda}

\def\t{\tau}
\def\s{\sigma}
\def\vs{\vskip .1 in}

\usepackage{varioref}
\usepackage{makeidx}
\makeindex

\usepackage[english]{babel}
\usepackage{caption}
\usepackage{subfig}

\numberwithin{equation}{section}

\usepackage{tabularx}
 
\begin{document}

\begin{spacing}{1.15}

\begin{titlepage}
\vspace*{2cm}
\begin{center}

{\Large \bfseries The Spectral Edge at Large Central Charge  \\ \vs and Black Hole Statistics}

\vspace*{1cm}

Eric Perlmutter

\vspace*{2mm}

\textit{\small Institut de Physique Th\'eorique, Universit\'e Paris-Saclay, CEA, CNRS\\Orme des Merisiers, 91191, Gif-sur-Yvette Cedex, France\\
\vskip .2 in
Institut des Hautes \'Etudes Scientifiques\\ 
35 Route de Chartres, 91440, Bures-sur-Yvette, France}

\vspace*{1cm}

\end{center}

\begin{abstract}

We introduce a new approach to high-energy spectra in AdS$_3$/CFT$_2$, addressing the open challenge of incorporating spectral statistics into the modular bootstrap. Using the analytic theory of $L$-functions attached to two-dimensional CFTs, we leverage modular constraints among high-energy states to derive a sum rule relating the spectral edge to bulk spectral fluctuations. As an application, we establish a refined modular bootstrap bound: letting $\Delta_1$ be the conformal dimension of the lightest excited Virasoro primary of a unitary CFT of central charge $c$, we show that $\Delta_1 - {c-1\o 12} \leq o(c)$, with at most one exceptional value of the scalar gap at each sufficiently large $c$. The result applies to theories with logarithmic rigidity of spectral counting fluctuations, as in random matrix theory. We exhibit circumstances under which the bound is strengthened further.

\end{abstract}

\end{titlepage}
\end{spacing}

\pagenumbering{roman}
\begin{spacing}{0.7}

\setcounter{tocdepth}{2}
\tableofcontents
\end{spacing}

\pagenumbering{arabic}
\setcounter{page}{1}

\begin{spacing}{1.07}


\sec{Introduction}\label{s1}

What is the largest possible spectral gap to the first non-vacuum Virasoro primary in a unitary, compact two-dimensional conformal field theory (CFT) of central charge $c$? This is the flagship question of the modular bootstrap program \c{Hellerman:2009bu}, which harnesses modular invariance of the torus partition function to constrain the local operator spectrum \c{Hellerman:2010qd, Keller:2012mr, Friedan:2013cba, Collier:2016cls, Afkhami-Jeddi:2019zci, Mukhametzhanov:2019pzy, Hartman:2019pcd,  Benjamin:2019stq}. See \c{Qualls:2013eha, Qualls:2014oea, Keller:2014xba, Benjamin:2016fhe, HKS, Dyer:2017rul, Collier:2017shs, Anous:2018hjh, Benjamin:2018kre, Ganguly:2019ksp, Pal:2019zzr, Gliozzi:2019ewk, Afkhami-Jeddi:2020hde, Mukhametzhanov:2020swe, BCM, 2107, Benjamin:2022pnx, DiUbaldo:2023hkc, Dey:2024nje, Pal:2025yvz, Benjamin:2026lbj, Erramilli:2026esf} for some work on variations of this theme. At large $c$, the best known results are $\D_1 \lesssim c/ 8.5$ analytically \c{Hartman:2019pcd}, and $\D_1 \lesssim c/9.1$ \c{Afkhami-Jeddi:2019zci} from numerical extrapolation, where $\D_1$ is the aforementioned gap. 

The modular bootstrap is a paragon of well-definedness. It is, at its root, a mathematical problem about non-holomorphic functions, with logical implications for unitary, modular-invariant CFTs. But this generality has a downside: it forecloses the possibility of {\it stratifying} the space of such CFTs. If one wants to probe the interior of theory space -- a core long-term goal of the non-perturbative bootstrap program -- one should filter out those CFTs with enhanced structure which obstruct algorithms that carve out this space.

This points to a logical tension in the folklore surrounding the modular bootstrap. For many years, there has been a latent consensus that the true upper bound on $\D_1$ asymptotes to $c/12$ as $c\rar\i$. The arguments for this primarily stemmed from holographic duality with AdS$_3$ gravity, and can be briefly summarized as the statement that every theory of AdS$_3$ gravity has BTZ black holes, which are locally AdS$_3$ geometries, starting at zero classical mass -- and thus, according to the basic AdS/CFT dictionary, the CFT side should possess dual primary operators at all conformal dimensions $\D \gtrsim c/12$ (up to $1/c$ corrections unimportant for this discussion). 

However, this rests on unexamined assumptions about the domain of validity of gravitational arguments. In particular, certain highly structured CFTs may not stand in correspondence with AdS$_3$ gravity as we know it. Indeed, a suggestion of the last several years is that geometric phenomena in semiclassical AdS$_3$ gravity should be understood not microscopically, but instead as certain {\it averaged} statements about CFT data. This immediately challenges the above gravity argument for a $\D_1 \lesssim c/12$ bootstrap bound: there can be microscopic CFTs to which gravity reasoning is simply not applicable. 

This point of view is supported by a pure CFT computation, in the nearby setting of Narain CFT. These are theories of $c$ free bosons compactified on particular lattices, which possess (two copies of) a $U(1)^c$ chiral algebra. It was proven in \c{Afkhami-Jeddi:2020ezh,Maloney:2020nni} that, with respect to the Zamolodchikov measure on Narain moduli space, the {\it typical} gap in the $U(1)^c$-primary spectrum is $\D_1 \approx c/2\pi e$ as $c\rar\i$. On the other hand, the best-known upper bound is significantly above this, around $c/9.797$ analytically \c{KabatyanskiiLevenshtein1978, CohnZhao2014} (where $2\pi e \simeq 17.079$). Moreover, the $c/2\pi e$ can be reproduced from a gravity-like computation. If one accepts this as a kind of proxy for the Virasoro case, then taken together with the previous paragraph, one has strong reason to scrutinize the conventional wisdom underlying modular bootstrap expectations. 

From the point of view of someone interested in AdS black hole physics, the question is more the converse: how do we bootstrap the CFTs that we really care about? In other words, how {\it do} we modify the modular bootstrap approach to be capable of stratifying the space of 2d CFTs? This amounts to understanding how to inject certain properties of the high-energy CFT data, in a manner which still effectively imposes modularity, unitarity and other desired constraints. This ``next-generation modular bootstrap'' is the central motivation of this paper. 

The black hole perspective offers a tip on what observable we should seek to incorporate: namely, {\it high-energy spectral statistics}. Black hole microstates in AdS gravity exhibit many strong signs of chaos and random matrix behavior \c{Shenker:2013pqa,cpp,MSS,Saad:2018bqo,CJ}. This dovetails with the independent, and much more general, lesson from quantum chaos that the semiclassical spectrum of any sufficiently non-fine-tuned system will exhibit random matrix fluctuation statistics \c{BGS}. The exceptions to this have a structural reason, such as extra arithmetic relations or symmetries. Spectral statistics, such as counting fluctuations and level correlations, are observables that may be defined unambiguously in a given deterministic system (no ensemble required), by suitably sampling the spectrum in energy bands. Indeed, it is our view that the phrase ``non-perturbative CFT'', usually used impressionistically, should be understood as a statement of the absence of non-generic structure in statistical measures of high-energy data. So if the ``old'' modular bootstrap question is to ask how large the gap can be in general (unitary, modular-invariant) CFTs, a ``new'' modular bootstrap may ask: {\it how large can the gap be in CFTs with non-trivial spectral statistics}? 

In this work, we take steps in this direction by combining {\it counting fluctuation statistics} with modularity and unitarity in a bootstrap approach. In particular, our results apply to the universality class of CFTs whose scalar spectra incorporate a hallmark of {\it random matrix statistics}. We are able to make progress by thinking about this problem from the perspective of $L$-functions attached to 2d CFTs \c{2509}. The basic construction of \c{2509} will be reprised in {\bf Section \ref{s2}}. The bootstrap setting is well-suited for this approach, in which the $L$-function $L_\cZ(s)$ is made from {\it differences} $\cZ(\t)$ of two putative partition functions: it is supported {\it only} on heavy states $\D > {c-1\o 12}$. As explained in Section \ref{s21}, taking a difference means that positivity is lost, but various features are gained. 

Before getting into any details, the main technical idea behind the $L$-function bootstrap can be explained as follows. $L$-functions admit a Hadamard product representation in terms of their zeros. When they also admit a convergent (generalized) Dirichlet series, these two representations share an overlapping regime of validity. One can thus, in the style of a non-trivial crossing equation, play them off each other: that is, we have an equation ``\textsf{Spectrum = Zeros}''. This gives a handle on some analytic structure of partition functions which is not visible in other approaches (and remains somewhat mysterious in conventional terms). 

\begin{figure}[t]
    \centering{\includegraphics[scale=.9]{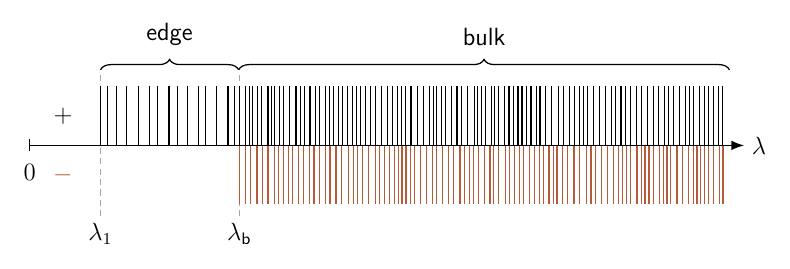}}
    \caption{We study differences of scalar spectra, which we split exactly into ``edge'' and ``bulk'' regions. In the bootstrap setting, we take $\lb \gtrsim c$ to be a fiducial gap, and upper bound $\l_1$.}
    \label{specfig}
\end{figure}

In {\bf Section \ref{s3}}, we will make this explicit by deriving sum rules that relate the scalar spectral gap to the series and the zeros, essentially obtained by taking the {\it logarithmic derivative} of the (completed) $L$-function. Letting $\l_1 \coloneq \D^{(0)}_1 - {c-1\o 12}$ be the shifted conformal dimension of the lightest non-vacuum scalar, we give an initial, heuristic explanation of why scalar gaps $\l_1$ scaling with $c$ are fine-tuned. 

{\bf Section \ref{s4}} then proceeds to a formal proof of a bootstrap bound using the sum rules. The series is where we inject the spectral statistical information into a ``bulk'' part of the spectrum, allowing an ``edge'' whose spectral statistics need not obey any particular constraint (see Figure \ref{specfig}). The zeros are subsequently treated using complex analysis, with {\it no} explicit assumptions about their distribution. 

The result, which is stated precisely in Section \ref{s4} (with useful background in Section \ref{s21}), can be summarized by the following schematic depiction of the {\it gap landscape} for theories with gaps $\D_1 > {c-1\o 12}$:

\begin{figure}[h]
    \centering{\includegraphics[scale=.81]{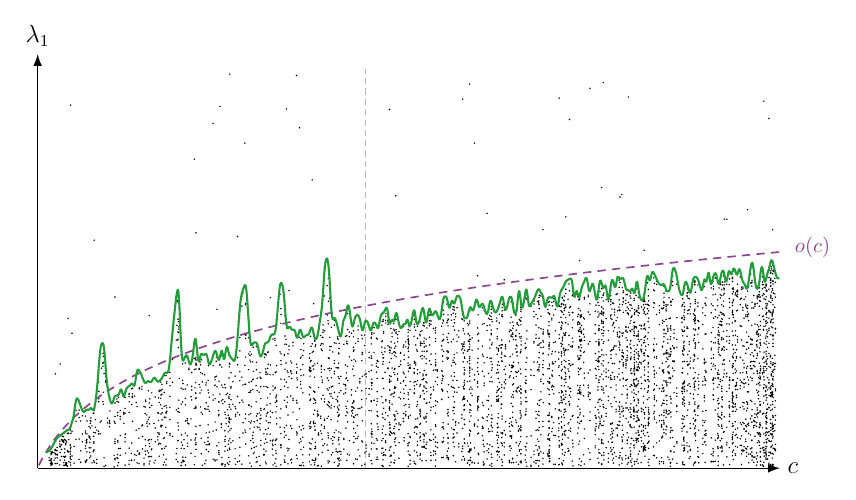}}
    \label{gapfig}
\end{figure}
\noindent Each point is a possible value of $\l_1 = \D^{(0)}_1 - {c-1\o 12}$. The green line represents the largest possible gaps: at sufficiently large $c$, they are upper-bounded by the dashed curve with vanishing slope, i.e. $\l_1 = o(c)$, modulo {\it one} possible exceptional gap, represented by the isolated points above the dashed curve. This is the result suggested by semiclassical AdS$_3$/CFT$_2$, and we are willing to tolerate the hypothetical exceptions (the likelihood of whose actual existence is left to the reader to judge). Moreover, there is even a yet {\it stronger} result (not pictured) for those exceptional points: namely, along a sequence of $c$ for which gaps of size $\gtrsim c$ are actually realized, the remaining gaps must obey $\l_1 = O(\log^2c)$, not merely $o(c)$. This arises because the actual proof proceeds by assuming that a putative gap $\lb \gtrsim c$ is possible, and then showing that all other gaps must be $O(\log^2 c)$; then, removing the assumption on $\lb$ leads to the $o(c)$ result. 

\begin{figure}[t]
    \centering{\includegraphics[scale=0.9]{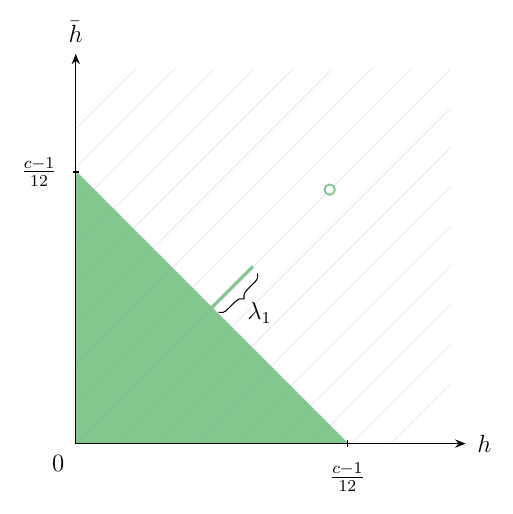}}
    \caption{At sufficiently large fixed $c$, there must be at least one non-vacuum Virasoro primary in the solid green region, modulo a single possible exceptional scalar gap denoted by the open circle. The line segment extends to $h+\bar h = {c-1\o 12} + \l_1$. In general, $\l_1 = o(c)$. Under further circumstances specified in the text, the bound becomes stronger. The gray lines remind us that spin is integer-quantized.}
    \label{specdetfig}
\end{figure}

Figure \ref{specdetfig} alternatively depicts the result as an allowed region for the gap in $(h,\hb)$ space. In fact, as explained in Section \ref{s21}, this makes clear that the method proves a more general result of {\it spectral determinacy} \c{kaidi, 2107}: at sufficiently large $c$, within the universality class of CFTs considered, the spectrum in the green region, whether empty or not, {\it uniquely fixes} the rest of the scalar spectrum. 

The proof is fully deterministic, but the underlying physical origin and justification for the bulk spectral assumption is statistical: it follows from {\it logarithmic rigidity of counting fluctuations}, a property inherited from random matrix statistics. We explain this sufficient condition in formal terms in {\bf Section \ref{s5}}, as well as the limitations of this assumption in pushing the bootstrap bound on $\l_1$ down even further below $o(c)$. This identifies a sufficient class of CFTs for which the bounds hold, in keeping with the goal of incorporating spectral statistics, here specialized to the generically-expected random matrix case, into a bootstrap bound. 

Looking ahead, while we have only bounded the spectral gap here, we anticipate that the sum rules from the $L$-function approach of Section \ref{s3} hold much more information about scalar spectra of 2d CFTs. In particular, they may be massaged into a functional approach for constraining higher levels, that can be used to probe the finer structure of heavy spectra. We leave this for future development. More broadly, a longer-term vision for the conformal bootstrap is to explore the high-energy, chaotic parts of the spectrum, rather than the {\it low-energy} states which are non-universal.

\sec{Setup}\label{s2}

Our object of study is the space of torus partition functions $Z(\t)$ of Virasoro CFTs -- that is to say, CFTs without Virasoro primary conserved currents -- of central charge $c>1$ and discrete spectra. Everything in this paper extends straightforwardly to CFTs with extended chiral algebras $\cA$ with $N$ generating currents using the results of \c{2509}, but we fix the notation to the $N=1$ Virasoro case both for universality and simplicity of presentation.

Consider a difference of putative Virasoro primary partition functions of central charge $c$:
\e{}{\cZ(\t) \coloneq Z_p(\t) - \tilde Z_p(\t)}
where 
\e{}{Z_p(\t) \coloneq \sqrt{y}|\eta(\t)|^2 Z(\t)}
and likewise for $\tilde Z_p(\t)$. The object $\cZ(\t)$ is a non-holomorphic modular function of $\t$: 
\e{}{\cZ(\g\t)=\cZ(\t)\,,\quad \g\in SL(2,\ZZ)}
We can thus restrict to $\t\in\cF \coloneq \{\t=x+iy: y>0, |x|\leq \half, |\t|\geq 1\}$, a fundamental domain of $SL(2,\ZZ)$ that includes the cusp at $\t=i\i$. As state sums, $Z_p(\t)$ and $\tilde Z_p(\t)$ count Virasoro primaries. By $T$-invariance, the local operator Hilbert space admits the integer-spin grading
\e{}{Z_p(\t)  = \sqrt{y} \sum_{j\in\ZZ}e^{2\pi i j x} Z_p^{(j)}(y)\,,\quad \text{where}~~Z_p^{(j)}(y) = \sum_{\D\geq |j|} d_{\D}^{(j)}e^{-2\pi \(\D-{c-1\o 12}\) y}}
and likewise for $\tilde Z_p(\t)$. In a unitary theory with $c>1$, the sums converge for $y>0$, and the spin-$j$ primary multiplicities $d_\D^{(j)}$ are non-negative for every allowed $(\D,j)$ (apart from the $j=\pm1$ vacuum module descendants, each of multiplicity $-1$). We restrict to unitary CFTs henceforth. 

For the problem of interest, suppose that $Z_p(\t)$ and $\tilde Z_p(\t)$ have identical spectra of ``light'' primaries, defined as those primaries with conformal dimensions $\D \coloneq h+\hb \leq {c-1\o 12}$. Then $\cZ(\t)$ is bounded as $y\rar\i$, and unitarity implies $\cZ(\t) \in L^2(\cF)$. Consequently, attached to this difference is an $L$-function, $L_\cZ(s)$, governed by a universal transformation law fixed by the modular and Virasoro symmetries \c{2509}: defining the completed $L$-function
\e{Lzdef}{\Lamz(s) \coloneq 2^s\g_{\cZ}(s) \lz(s)\,,}
with gamma factor
\e{gcft}{\g_\cZ(s) = \pi^{-2s} \prod_{i=1}^4 \G\({s+\k_i\o 2}\)\,,\quad \{\k_i\} = \{0,1,-\thalf,\thalf\}\,,}
$\Lamz(s)$ satisfies the functional equation of a self-dual $L$-function of degree 4 and root number $\vareps=1$,
\e{LZFE}{\Lamz(s)=\Lamz(1-s)\,.}
The functional equation is an avatar of modular invariance. $L_\cZ(s)$ is a Rankin-Selberg $L$-function, constructed from a modular integral of $\cZ(\t)$ against an $SL(2,\ZZ)$ Eisenstein series, which projects onto the scalar sector of $\cZ(\t)$. 

By insisting that their light spectra match exactly, every state in the cuspidal expansion of $\cZ(\t)$ is a heavy CFT state, with a sign grading enforced by unitarity: positive degeneracies have support on $Z_p(\t)$, and negative degeneracies have support on $\tilde Z_p(\t)$. All states are physical states.\foot{As explained in \c{2509}, each individual partition function may actually be attached to an $L$-function with the above transformation law, constructed by trading their light states for a suitable modular complement of heavy states that altogether preserve modularity. The viewpoint of this paper is different: $\cZ(\t)$ is the difference of two {\it bona fide} partition functions, each of which corresponds to a putative unitary CFT. There is no modular completion needed.}  So the $L$-function $L_\cZ(s)$ captures the discrete spectra of heavy scalar primaries of this pair of CFTs. 

The strategy suggested by this formulation for bootstrapping the spectral gap may already be evident, and indeed was presaged in \c{kaidi}, though at this point it is probably unclear what advantage may be afforded by the $L$-function presentation. Proceeding in that direction, we now review two representations of the $L$-functions. 

The first representation is a product over non-trivial zeros $\{\rho\}$,
\e{had}{\Xi(s) \coloneq s(1-s)\L_\cZ(s) = e^A \prod_{\rho}\(1-{s\o\rho}\)\,,}
where 
\e{}{\L_\cZ(\rho) = \L_\cZ(\rhob) = \L_\cZ(1-\rho) = \L_\cZ(1-\rhob) = 0\,.}
Zeros $\{\rho,1-\rho\}$ are understood as paired in the product. The constant $e^A$ will drop out of our computations. The $\ZZ_2 \x \ZZ_2$ ``quartet'' structure of the zeros follows from the self-dual functional equation. 

The second representation is a series. We have
\e{Z0y}{\int_{-\half}^\half \cZ(x+iy) dx = \sqrt{y} \sum_{\l>0}a_\l e^{-2\pi \l y}\,,}
where 
\e{}{\l \coloneq \D^{(0)}-{c-1\o 12}}
is a shifted scalar conformal weight and $a_\l \coloneq d_\D^{(0)}-\tilde d_\D^{(0)}$ is a difference of state multiplicities. The $L$-series is then
\e{zzdef}{\lz(s) = \z(2s)\sum_{\l>0} {a_\l\o \l^{s-\half}}\,,\quad a_\l\in\Z\,.}
The explicit series on the right-hand side is the spectral zeta function attached to $\cZ^{(0)}(y)$ -- a generalized Dirichlet series over the {\it scalar data} of $\cZ(\t)$ \c{2107, 2509}. We choose the sign convention $a_{\l_1}>0$: by unitarity, the leading frequency of $\cZ(\t) = Z_p(\t) - \tilde Z_p(\t)$ must have support on $Z_p(\t)$. The topic of series convergence will be addressed below. 

It will be useful to define the renormalized object
\e{Adef}{A(s) \coloneq \l_1^{s-\half}{L_\cZ(s)\o \z(2s)}}
In the series representation, stripping off the leading term will allow us to separate the structure of the series from the overall normalization. In terms of renormalized frequencies $\g$,
\e{Aseriesrep}{A(s) = \sum_{\g\geq 1}{a_\g\o \g^{s-\half}}\,,\quad \g \coloneq {\l\o \l_1} \geq 1}
where $a_1 \in\Z_+$. The completed $L$-function may be written in terms of $A(s)$ as
\e{lama}{\Xi(s) = G(s)\l_1^{\half-s} A(s)\,,\quad \text{where} \quad G(s) \coloneq s(1-s)2^s\g_{\cZ}(s)\z(2s)}
Henceforth, $\{\rho\}$ denotes the zeros of $\Xi(s)$, counted with multiplicity. For $\Re(s)>1$, where $G(s)$ is holomorphic and nonvanishing, these coincide with the zeros of $A(s)$.

\ssec{Bootstrapping with $\cZ(\t)$}\label{s21}

Certain features are gained relative to standard bootstrap approaches, despite having lost positivity in forming the difference $\cZ(\t)$:

\begin{enumerate}[label*=  \sf (\roman*)]

\item There is no vacuum module, and hence no asymptotic averaged Cardy background. 

\item Modular invariance acts only among high-energy states.

\item Modular invariance can be refined to act only among scalars.\foot{See \c{2107, Benjamin:2022pnx, Tierz:2025xhp, Benjamin:2025kvm} for earlier work on modular constraints for scalars. In \c{Tierz:2025xhp}, an argument was given that the scalar gap cannot exceed $c/12+O(1)$ as $c\rar\i$. However, the argument applies only to putative unitary, compact CFTs whose non-vacuum module primaries are {\it all} scalars -- a condition which is forbidden by the Cardy formula \c{Cardy:1986ie,Mukhametzhanov:2020swe}.}

\end{enumerate}

\noindent Moreover, it is important to stress that relics of unitarity {\it are} present, in two ways: first, by guaranteeing that $\cZ(\t) \in L^2(\cF)$, a necessary condition for the Rankin-Selberg construction of $L_\cZ(s)$; and second, by sign-grading the $L$-series \eqr{Aseriesrep}. In particular, unitarity implies that every frequency $\l$ for which $a_\l>0$ has support on $Z_p(\t)$, and likewise for $a_\l<0$ and $\widetilde Z_p(\t)$. 

Let us explain how to interpret bootstrap bounds on $\cZ(\t)$. Let $\{\l_n\}$ be the unbounded increasing sequence of positive scalar frequencies appearing in our CFT partition functions, without multiplicity: $0 < \l_1 < \l_2 < \ldots$, with $\l_{n\rar\i} \rar\i$. Let $\L(c)$ be the set of allowed values of $\l_1 \coloneq \inf\{\l_n\}$ at central charge $c$, perhaps for some subclass of CFTs. An upper bound derived from the difference $\cZ(\t)$ is necessarily a pairwise supremum bound, of the form
\e{}{\sup_{\l\neq \l' \in \L(c)} \min\{\l,\l'\} < M}
for some $M>0$. With respect to the large $c$ limit, a uniform pairwise $O(1)$ bound, for example, would be the statement that 
\e{limsupbound}{\limsup_{c\rar\i}\(\sup_{\l\neq \l' \in \L(c)} \min\{\l,\l'\}\)<\i}
This is mathematically equivalent to the formulation
\e{}{\exists M<\i\,,\exists \,c_0<\i~\text{s.t.}~\forall\, c\geq c_0\,,~~ \#\{\l\in\L(c):\l>M\} \leq 1}
There is at most one exceptional gap value.

Now, since $\cZ(\t)$ is formed by first matching light spectra, any bounds arising from $\cZ(\t)$ apply universally to pairs of partition functions in every light-spectral class. To bound the largest possible spectral gaps, we choose the sparsest possible class: namely, that for which the set of non-vacuum module primaries with $\D \leq {c-1\o 12}$ is empty. The bootstrap conclusion implied by the above bounds, therefore, would be that at least one of the following is true: 

\begin{enumerate}[label*=  \sf (\roman*)]

\item There is a light primary, of whatever spin, with $\D \leq {c-1\o 12}$.

\item There is a scalar primary with $\D \leq {c-1\o 12}+M$, modulo one possible exceptional gap. 

\end{enumerate}

\noindent Combining these implies the unified statement: over {\it all} spins, the lightest non-vacuum primary must obey $\D \leq {c-1\o 12} + M$, except possibly one exceptional scalar gap. If this approach bears fruit, a possible exceptional gap sequence is certainly something one is willing to live with. Of course, depending on one's priors, such exceptions may reasonably seem non-existent, but in the bootstrap spirit, we remain agnostic. 

The construction of $\cZ(\t)$ as a difference implies that a result like \eqr{limsupbound} should really be fundamentally interpreted as a form of {\it spectral determinacy} \c{kaidi, 2107}: for two partition functions in {\it any} fixed light-spectral class, the first heavy scalar dimension must obey the bound derived above. This is depicted in Figure \ref{specdetfig} (with $M$ written as $\l_1$ there). 

The next section explains how the $L$-function attached to $\cZ(\t)$ can actually be used to realize the potential of the above approach. 

\sec{Sum rules for the spectral gap}\label{s3}

The key idea in using $L$-functions to bootstrap CFT spectra is to use their analytic structure, playing the zeros and the spectrum off each other. 

\ssec{Log derivative}
Inspired by similar ideas in the study of zeros of standard $L$-functions, we can generate a sum rule for $\l_1$ by taking the logarithmic derivative of $\Xi(s)$. Using \eqr{had} and \eqr{lama} leads to the following equation:
\e{ldsumrule}{\boxed{\log\l_1 = F(s) + {A'\o A}(s) - \sum_\rho{1\o s-\rho}}}
where
\e{Fdef}{F(s) \coloneq 2 \psi(2 s-1)-\log 8 \pi ^2+2{\zeta'\o \z}(2 s)+{1\o s}+{1\o s-1}\,.}
The zeros $\{\rho\}$ are the non-trivial zeros of $\L_\cZ(s)$, which are also the non-trivial zeros of $L_\cZ(s)$, and hence of $A(s)$ (modulo the $\z(2s)$ factor in \eqr{Adef}). Note that $\log\l_1 \in \RR$ by unitarity. 

As stated, \eqr{ldsumrule} is simply an identity for the log derivative of $A(s)$ in the Hadamard representation. However, the key point is as follows: if we insert the {\it series} representation of $A(s)$ into \eqr{ldsumrule} in circumstances where it converges, the overlapping regions of convergence imply mutual constraints on the respective data, and in turn on $\l_1$. In particular, the series and zero data depend implicitly on $c$, and are non-trivially related to one another -- and to $\l_1$ -- by the above equation. This suggests that, barring mutual cancellations, large $\l_1$ may be difficult to achieve for a sufficiently random or structureless spectrum and zero set.

To reinforce this point, we note that the self-dual functional equation allows us to write the zeros in terms of a ``quartet kernel'' $K_q(s,\rho)$, whereupon \eqr{ldsumrule} becomes
\e{sum1}{\log\l_1 = \Re\(F(s) + {A'\o A}(s)\) + Z(s)}
with 
\es{}{Z(s) &\coloneq -\half\sum_{\rho\,:\,\Re(\rho)\geq \half}\Re(K_q(s,\rho))\\
K_q(s,\rho) &\coloneq {1\o 1+\d_{\Re(\rho),\half}}\left({1\o s-\rho} + {1\o s-\rhob}+{1\o s-(1-\rho)}+{1\o s-(1-\rhob)}\right)}
The zero term $Z(s)$ is a sum only over zeros on or to the right of the critical line, $\s=\half$. The prefactor avoids double-counting critical zeros. From the explicit form of the kernel, one can deduce a key conceptual fact: {\it for a fixed $s$, only zeros with $\Re(\rho) > \Re(s)$ contribute positively to $Z(s)$}. This is powerful because, given a right half-plane $\Re(s) > \s_*$ in which the series for $A(s)$ converges, all of the ``bad'' zeros live in that same half-plane. Critical zeros, if present, only reduce the right-hand side of \eqr{sum1}. This reinforces the spectrum-zeros interplay. 

Suppose, then, that $\l_1 \sim \b c^\a$ as $c\rar\i$ for some $\a,\b\in\R_+$, so that
\e{scaling}{\a\log c \approx \Re\(F(s) + {A'\o A}(s)- \sum_\rho{1\o s-\rho}\)}
Our primary goal is to ask whether $\a=1$ is allowed, and more broadly, whether any $\a>0$ is allowed. Below we give an initial, heuristic treatment of this point, before proceeding to the proof in the next section.

\ssec{No large gaps: Heuristics} 

To warm up, let us understand, in qualitative terms, what $\log c$ growth of either term on the right-hand side of \eqr{scaling} would entail. We explain the physical sense in which the necessary mathematical events are ``extreme'', i.e. fine-tuned. 

To set things up, we first point out that $\log\l_1$ is (obviously) $s$-independent. The $c$-independent ``kinematic'' term $F(s)$ obeys 
\e{Fasymp}{\qquad\qquad\qquad\qquad\Re(F(s)) = \log |s|^2 + O(1)\qquad (1<h<\Re(s))}
with $O(1)$ constant depending implicitly on $h$. This suggests that we analyze whether there can exist points $s$ with $|s|^2 = o(c)$ at which either of the series or zeros terms can give a logarithmically divergent contribution to \eqr{sum1} as $c\rar\i$. 

\sssec{Series}

We start with the series term. Assume we work at a value of $s$ where the series converges, and hence
\e{}{A'(s) = -\sum_{\g>1} {a_\g \o \g^{s-\half}}\log \g}
Let us explain why ${A'\o A}(s)\sim \log c$ scaling requires fine-tuning. Since the $\g=1$ term does not participate in the derivative, a necessary condition for a large log derivative is that the subleading terms $\g>1$ have significant variation in $s$ at large $c$, for every $s$. Indeed, the condition we are exploring effectively amounts to
\e{Aeig}{\qquad \qquad ~~A(s) \approx c^s f(s) \qquad (c\rar\i)}
for $f(s)$ slowly varying, for {\it all} $s$ in the half-plane of convergence.\foot{We emphasize that ${A'\o A}(s)\sim \log c$ scaling is not difficult to engineer at isolated $s$ with sufficient fine-tuning among the data $\{\l,a_\l\}$; the challenge is that it must hold {\it uniformly in $s$ in a half-plane of convergence}, whereas the data $\{\l,a_\l\}$ are {\it $s$-independent}.} To generate such exceptional behavior, a first reflex might be to look for a saddle point at some $\g\approx \g_*(c)$ scaling with $c$ which would give ${A'\o A}(s)\sim \log c$ growth. However, no such saddle can exist in the physical range of summation: a saddle at $\g_*>1$ would yield a {\it negative} enhancement to the log derivative,
\e{}{{A'\o A}(s) \approx -\log\g_* <0}
rather than the positive $\log c$ growth we are after. The necessary enhancement must instead involve delicate cancellations among spectral contributions. For instance, an approximate lattice structure among the frequencies, or multiplicative structure of the coefficients, might give rise to this emergent scaling behavior, but anything approaching a generic spectrum of many weakly-correlated states would not. The proof to follow controls these cancellations through a counting bound and a comparison condition, while treating the zero contribution (discussed below) by complex analysis.

\sssec{Zeros}

Can the zero term $Z(s)$ generate a $\log c$ growth? Asymptotically along rays, 
\e{sasymp}{\Re(K_q(s,|\rho|e^{i\theta_\rho})) \approx {2(1-2\s)\cos(2\theta_\rho)\o |\rho|^2} \qquad (|\rho|\rar\i\,,~\theta_\rho~\text{fixed})}
$\Xi(s)$ is an order one function, which implies 
\e{}{\sum_\rho {1\o |\rho|^{1+\eps}}<\i~~\forall\,\eps>0}
Relative to the naive $|\rho|^{-1}$ scaling expected from \eqr{ldsumrule}, \eqr{sasymp} enjoys an extra $|\rho|^{-1}$ savings as a consequence of self-duality of $L_\cZ(s)$.

\begin{figure}[t]
    \centering{\includegraphics[scale=1]{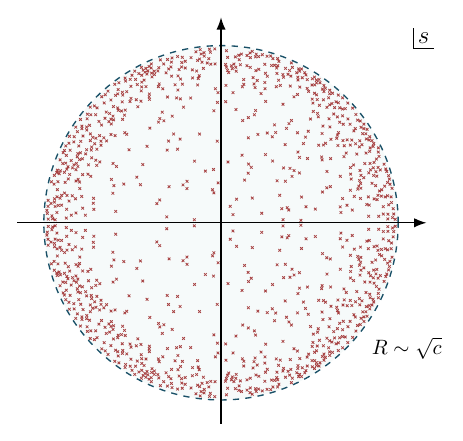}}
    \caption{A schematic depiction of a hypothetical ``dense disc'' of zeros, shown here accumulating inside a disc of radius $R \sim \sqrt{c}$ as $c\rar\i$.}
    \label{zerofig}
\end{figure}

Assume there is a point $s_*$ which is uniformly bounded away from all zeros as $c\rar\i$. Then as $c\rar\i$, the scaling $Z(s_*) \sim \log c$ could only arise if the {\it zero density} diverges with $c$. We call this a ``dense disc'' scenario, in which zero sparsity bounds are violated by divergent accumulation of zeros at large $c$. We can be a bit more precise by using the specific structure of $A(s)$. Let
\e{}{N_\rho(R) = \#\left\{\rho : \Big|\rho-\half\Big|\leq R\,,~\Xi(\rho)=0\right\}}
count zeros inside a disc centered at $s=1/2$. Jensen's formula implies the general scaling $N_\rho(R) = O(R^{1+\eps})~\forall\, \eps>0$. The quartet symmetry then gives rise to a natural ``zero sparsity condition'' 
\e{zscond}{\int_1^\infty\frac{N_\rho(R)}{R^3}\,dR=O(1)}
But in principle, this may be violated as $c\rar\i$ -- for example, the error term from Jensen's formula may grow with $c$, such that it competes with the leading term for points inside some $c$-dependent radius. See Figure \ref{zerofig}. 
 
As a point of contrast, we recall the famous fact that standard $L$-functions have, unconditionally, {\it no} non-trivial zeros outside the critical strip \c{IK}. If this were the case for CFT $L$-functions $L_\cZ(s)$ as well, the zeros would be harmless for the above purposes, because they would contribute negatively to $Z(s)$ (for $\Re(s)>1$). 

A closer point of comparison is to $c\leq 1$ Virasoro CFTs, whose $L$-functions we can construct explicitly. Neither $c=1$ free boson CFTs, nor $c<1$ Virasoro minimal models, have zeros outside the critical strip. Moreover, one could ask whether the zeros approach the boundary of the critical strip as one takes $c\rar 1$ through a sequence of minimal models, which would plausibly suggest the existence of zeros with $\Re(\rho)>1$ for Virasoro CFTs at $c>1$. However, the zeros do not exhibit this behavior, instead condensing along the critical line.\foot{We thank Yanjun Zhou for discussions that led to this observation.} 

\ssec{Blaschke bounds}
The sum rule \eqr{ldsumrule} may be massaged to yield a powerful set of inequalities. They take advantage of a property similar to the zero-sign observation made below \eqr{sum1}. 

Define the half-plane maximum
\e{Qmax}{Q(h)\coloneq\sup_{\Re(s)\ge h}\frac{|A(s)|}{|s+1|}.}
The relevant positive range of $h>1$, lying inside a half-plane of convergence of $A(s)$, will be specified later. We will consider points $s_0$ in the half-plane $\Re(s)> h$, 
\e{}{s_0=\sigma_0+it_0\,,\quad d\coloneq\sigma_0-h>0\,.} 
There will be three bounds:

\begin{itemize}
\item The first bound, which holds for all such points $s_0$ with $A(s_0)\neq0$, is
\e{bweak}{\boxed{ \log\lambda_1\le
 \Re F(s_0)+\Re \frac1{s_0+1}
 +\frac1d\log\frac{Q(h)|s_0+1|}{|A(s_0)|}}}
$A(s)/(s+1)$ is analytic for $\Re(s)>h$, and convergence of $A(s)$ implies that $A(s)/(s+1)$ tends uniformly to zero as $|s|\to\infty$ in $\Re(s)\geq h$; so its positive supremum is attained at a finite point, which the maximum principle places on the boundary, $\Re(s)=h$. 

\item A stronger bound exists when $s_0$ sits at this boundary maximum point, call it $s_* \coloneq h+it_*$, defined by
\e{bmax}{{|A(h+it_*)|\o |h+1+it_*|} \coloneq Q(h)}
with $t_*\in\R$. In this case,
\e{bstrong}{\boxed{\log\l_1 \leq \Re F(h+it_*) + \Re{1\o h+1+it_*}}}

\item For any boundary point $s_0=h+it_0$ with $A(s_0)\neq0$, a modification of the same argument gives the following bound, which will be most useful for our purposes:
\e{logbound}{\boxed{\l_1 \leq C \[h+|t_0|+\log\(1+{Q(h)\o |A(s_0)|}\)\]^2}}
where $C$ is a uniformly bounded constant. 

\end{itemize}

\sssec{Proof}

The proof of all of these statements follows from representing the zeros using a Blaschke product. A Blaschke product is designed to reflect zeros in a chosen half-plane. For every zero with $\Re(\rho)>h$, define
\e{}{\rho_h^* \coloneq 2h-\rhob\,,\qquad b_\rho(s) \coloneq \eta_\rho {s-\rho\o s-\rho_h^*}\,,\qquad |\eta_\rho|=1}
The Blaschke product is
\e{}{\cB_h(s) \coloneq \prod_{\Re(\rho) > h} b_\rho(s)}
where a zero $\rho$ and its conjugate $\rhob$ are both included here (and phases $\eta_\rho$ chosen to ensure its convergence). It is manifest that $|\cB_h(s)|\leq1$ for $\Re(s) > h$, with saturation on the boundary. The key point is that dividing the Hadamard product by the Blaschke product $\cB_h(s)$ reflects zeros with $\Re(\rho)>h$ across the line $\Re(s)=h$.

We can factor our function $A(s)$ into a Blaschke product and a bounded remainder as follows:
\e{Abla}{{A(s)\o Q(h)(s+1)} = \cB_h(s)g_h(s)\,,\quad |g_h(s)|\leq 1\,,\quad g_h(s)\neq 0 \quad (\Re(s)>h)}
Blaschke factorization implies that $g_h(s)$ is bounded, analytic and zero-free for $\Re(s)>h$. Plugging the representation \eqr{Abla} into \eqr{lama},
\e{xib}{{\Xi(s)\o \cB_h(s)} = {\l_1^{\half-s}Q(h)(s+1)G(s)}g_h(s)}
This is the ratio of the Hadamard and Blaschke products. The fact that this ratio only has zeros that lie {\it outside} the half-plane $\Re(s) > h$ leads to a useful positivity constraint on the log-derivative of the ratio. In particular, at a point $s_0$ with $\Re(s_0) \geq h$ and $A(s_0)\neq 0$,
\e{rexibd}{\Re \({\Xi'\o \Xi} - {\cB_h'\o\cB_h}\)(s_0) = \sum_{\Re(\rho)\leq h}\Re{1\o s_0-\rho} + \sum_{\Re(\rho)>h}\Re{1\o s_0-\rho^*_h} > 0}
In particular, every zero on the right-hand side has real part at most $h\leq\Re(s_0)$, so each term is non-negative (and the $\rho \rar 1-\rho$ symmetry implies at least one nonzero term).\foot{This rearrangement of zeros in the log-derivative is legitimized by boundedness and the functional equation for our order one function (cf. Section \ref{s2}), which imply the Blaschke condition
\e{}{\sum_{\Re(\rho)>h} {\Re(\rho)-h\o 1+|\rho|^2} <\i}
}
Therefore, taking the (real part of the) log derivative of both sides of \eqr{xib}, and using \eqr{rexibd}, yields a bound: after rearrangement,
\e{rearrange}{\log\l_1 \leq \Re \(F(s_0) + {1\o s_0+1}+{g'_h\o g_h}(s_0)\)}

Let us first evaluate this at the boundary maximum point $s_0 = s_*$. In that case, the left-hand side of \eqr{Abla} has unit modulus, and $|\cB_h(s_*)|=1$, so $|g_h(s_*)|=1$. Moving horizontally into the half-plane cannot increase this modulus, so $\partial_\s\log|g_h(\s+it_*)||_{\s=h}\leq 0$. Since the horizontal derivative is the real part of the logarithmic derivative, we arrive at the following properties of $g_h(s)$ at the boundary maximizing point:
\e{gbdy}{|g_h(s_*)|=1\,,\quad \Re\( {g'_h(s_*)\o g_h(s_*)} \)\leq 0}
Plugging \eqr{gbdy} into \eqr{rearrange} implies the bound \eqr{bstrong}.

Closely related logic leads to the bound \eqr{bweak}, valid for any point $s_0$ with $\Re(s_0)>h$. Because $g_h(s)$ is analytic and zero-free for $\Re(s)>h$, $v_h(s_0) \coloneq -\log |g_h(s_0)|$ is a non-negative harmonic function, which implies $|\partial_\s v_h(s_0)| \leq v_h(s_0)/d$. Since $|\cB_h(s_0)|\leq 1$ implies, from \eqr{Abla},
\e{}{v_h(s_0) \leq \log{Q(h)|s_0+1|\o |A(s_0)|}\,,}
plugging into \eqr{rearrange} yields \eqr{bweak}. 

Finally, let us derive \eqr{logbound}. The technique is similar to the above, only we consider a different norm for $A(s)$. Define, for $k\in\Z_{\geq 2}$,
\e{}{f_k(s) \coloneq {A(s)\o (s+k)^k}\,,}
which obeys
\e{}{|f_k(s)| \leq Q(h) {|s+1|\o |s+k|^k}\,.}
All of the analytic conditions used previously also hold here. In particular, the right-hand side again tends to zero at infinity in the half-plane $\Re(s)>h$, and by the maximum principle, the maximizer of $|f_k|$, call it $s_{*,k} \coloneq h + it_k$, is again on the boundary. Similar steps as those which led to \eqr{bstrong} yield its $k$-fold analog,
\es{logboundint}{\log\l_1 &\leq \Re F(h+it_k) + k \Re{1\o h+k+it_k} \\&\leq \Re F(h+it_k) + 1}
The point is that we can now take advantage of $k$ to locate the maximum. Let us take our point $s_0 = h + it_0$ on the boundary. For a given such point, if $k\geq h+|t_0|$, then
\e{}{ {|A(s_0)|\o (2k)^k} \leq |f_k(s_0)|\,.}
At other points on the boundary for which $|t|\geq 8k$, say, we have 
\e{}{|f_k(h+it)| \leq 2Q(h) |t|^{1-k}\,.}
Comparing these two bounds, 
\e{}{{|f_k(h+it)|\o |f_k(s_0)|} \leq 16 {Q(h)\o |A(s_0)|}k 4^{-k}}
If we choose $k$ sufficiently large in terms of $Q(h)/|A(s_0)|$, then the right-hand side is less than one: therefore, the maximizer must live {\it inside} the band 
\e{}{|t_k| < 8k}
for the chosen $k$. This is, in particular, a logarithmic upper bound in $Q(h)/|A(s_0)|$. Plugging the asymptotics \eqr{Fasymp} into \eqr{logboundint}, and using the upper bound on the maximizer $t_k$ just derived, gives \eqr{logbound}, with uniform constant $C$.

\sec{Spectral bound}\label{s4}

In this section we will turn the bounds on $\log\l_1$ derived in Section \ref{s3} into bootstrap bounds. 

The starting point is the renormalized series $A(s)$. We split this into what we call ``edge'' and ``bulk'' terms:
\e{Adeff}{A(s) = A_{\edge}(s) + A_{\bulk}(s)\,, \quad A_\edge(s) \coloneq \sum_{\g<\gb} {a_{\g}\o \g^{s-\half}}\,,\quad \gb = {\lb\o\l_1}>1}
The frequency $\gb$ is defined as the first frequency for which $a_\g<0$: 
\e{}{a_1\geq 1\,,\qquad a_\g\geq 0~~\forall\,\g\in(1,\gb)\,.}
See Figure \ref{specfig}. Thus the edge is a sum over positive multiplicities, and so 
\e{edgepos}{|A_\edge(\s+it)| \leq A_\edge(\s)\,,\quad A_\edge(\s)\geq 1}
$\lb$ is the ``bulk threshold'', interpreted as the frequency where the two spectra start to mix: absent exact degeneracies, $A_\edge(s)$ is comprised of states only belonging to $Z_p(\t)$. The edge/bulk decomposition of $A(s)$ is exact. It will be useful to further (exactly) split the bulk into two pieces,
\e{Abulkdeff}{A_\bulk(s) = \bar{A}_\bulk(s) + \d A_\bulk(s)}
$\bar{A}_\bulk(s)$ is defined with respect to a reference density, while $\d A_\bulk(s)$ is the fluctuation. 

We will use the different physics of these regions in the proof in the following way. First, the bulk region is where a counting fluctuation bound injects statistical universality of the partition functions. This in turn controls the size of $\d A_\bulk(s)$, in a precise manner to be seen below. Combined with positivity of the edge coefficients, a mild non-cancellation condition between the bulk and edge required on a single vertical line (which rules out extreme fine-tuning) and the bound \eqr{bstrong} from the analytic structure of the $L$-function will give a bound on $\l_1$. The bound does not require the reference densities of $Z_p(\t)$ and $\tilde Z_p(\t)$ to coincide, nor bounds on their absolute size.

Our initial goal is to prove $\l_1 = o(c)$, strengthening in due course. Since $\l_1 < \lb$, we consider a fiducial partition function with $\lb \gtrsim c$. We impose no {\it upper} bound on $\lb$, i.e. we are not making any assumption about how large the fiducial gap can be.

\ssec{Assumptions and statement}
Let us first state our assumptions, whose physical content we explain after the proof. Unless otherwise noted, all constants are $c$-independent. 

Our first assumption is a deterministic bound on the bulk counting fluctuations, which we state as follows. Let $\nu(I)$ be the signed state count over a bulk interval $I \subseteq [\lb,\i)$, which by \eqr{Abulkdeff} admits a decomposition $\nu(I) = \bar\nu(I) + \d\nu(I)$. Let $S(\l) \coloneq \nu([\lb,\l))$ be the cumulative count up to some $\l>\lb$, which also admits a decomposition $S(\l) = \bar S(\l) + \d S(\l)$. The condition is that, for $\lb\gtrsim c$,
\e{Sass}{\exists\,a,b\geq 0~~\text{s.t.}~~|\d S(\l)| \lesssim c^b \l^a\,.}
Any fixed values of $(a,b)$ will suffice (though we note that \c{HKS} implies $a+b\leq 2$). In fact, the proof really uses the Abel summation consequence \eqr{abel}, for which \eqr{Sass} is just a sufficient condition with a clear physical origin. We will later explain this as a property of a sum over bulk intervals whose statistical counting fluctuations obey {\it log-rigidity}.

The second assumption is a non-cancellation condition between the edge and bulk. We assume that there exists at least one vertical segment $H+it$ with $t\in [-T,T]$ and $T\geq 0$, with fixed $H$ in right half-planes specified below, where 
\begin{equation}
 \frac{A_{\edge}(H)}
 {\sup\limits_{|t|\le T}|A(H+it)|} \lesssim c^p
 \label{eq:comparison}
\end{equation}
for some fixed $p\geq0$. This extremely mild assumption asserts that the spectral pairs in question are not highly fine-tuned as $c\rar\i$: by requiring only one such segment, we are forbidding that,  along vertical lines {\it in an entire right half-plane}, the bulk cancels the edge to precision greater than any power of $c$ relative to $A_{\edge}(H)$. (In Section \ref{s421}, $H>h_0+p+2, h_0\geq p+{5\o2}$, while in Section \ref{s424}, $H> \max\{1,a+\half\}$.)

We will use these assumptions to prove spectral bounds, with a few variations. In the first proof, in which we set $\bar A_\bulk(s)=0$, these assumptions will imply $\lambda_1=O(1)$ when $\lb \gtrsim c$; removing the assumption $\lb \gtrsim c$ further implies the general result $\l_1 = o(c)$. We then extend the $\l_1 = o(c)$ result to $\bar A_\bulk(s) \neq 0$.\foot{The mechanics of the proof do not require $\lb \gtrsim c$, but so as not to overextend our assumptions, we only take \eqr{Sass} to hold in that range. Consequently, the bounds are of the form $\l_1 = o(c)$, even when the formal proof appears to do better. We will discuss the physical implications of a would-be extension to $\lb = o(c)$ in Section \ref{s5}.}

\ssec{Proof}
The general proof is in Section \ref{s424}. We begin in Section \ref{s421} by setting $\bar A_\bulk=0$, relaxing this later on. This is a useful starting point for a few reasons: as an instructive warmup; as a bound applicable to theories which have universal Cardy asymptotics for all $\l\geq \lb$ (to be explained in Section \ref{s53}); and because it leads to a slightly stronger bound than the general case. 

\sssec{Warmup: Bounded bulk density}\label{s421}
The basic strategy of our initial proof is to use \eqr{bstrong}. For fixed $h$, the right-hand side of \eqr{bstrong} can only grow large for large $t_*$: using \eqr{Fasymp},
\e{l1t}{\l_1 \leq C_h(1+t_*^2)}
as $t_*\rar\i$ with bounded $C_h$. Thus, an upper bound on $t_*$ yields an upper bound on $\l_1$. In particular, $t_* = o(\sqrt{c})$ would imply $\l_1 = o(c)$, while $t_* = O(1)$ would imply $\l_1 = O(1)$. Our analytic proof will amount to the following physical statement: {\it if the bulk obeys log-rigid counting statistics around a fixed reference density, then $t_* \rar 0$}. 

To reduce the number of variables flying around we here specialize to 
\e{abvals}{a={3\o2}\,,\quad b=\half\,.}
General $(a,b)$ may be restored in the proof (and will be in Section \ref{s424}) at the cost of adjusting the fixed auxiliary parameters and convergence half-planes.\foot{As we will explain, these are the ``natural'' values associated to the Cardy formula.} The allowed range of $H$ is 
\e{hrange}{h_0\geq p+{5\o2}\,,\quad  H = h_0+d\,,\quad d>p+2}
We emphasize that no specific value of $H$ or $d$ within this range is singled out in the computation below.

There are essentially four steps to deriving $\l_1 = O(1)$.

As a first preliminary step, by Abel summation,
\begin{equation}
 A_{\bulk}(s)
 =(s-\tfrac12)\lambda_1^{s-1/2}
 \int_{\lb}^\infty S(\l)\l^{-s-1/2}\,d\l,
 \qquad \Re(s)>2.
 \label{eq:abel}
\end{equation}
The boundary term at infinity vanishes. Define the {\it bulk} maximum in any half-plane, analogous to \eqr{Qmax},
\e{}{Q_{\bulk}(h)\coloneq\sup_{\Re(s)\ge h}\frac{|A_{\bulk}(s)|}{|s+1|}}
Since $|s-1/2|/|s+1|\le1$, plugging \eqr{Sass} and \eqr{abvals} into Abel gives, uniformly over the entire half-plane,
\begin{equation}
Q_{\bulk}(h) \lesssim \sqrt{c}\lambda_1^{3/2}\gb^{2-h},
 \qquad h>2.
 \label{abel}
\end{equation}

Next, we take our point $s=H+it$ to be the point $s_0$ defined by $|A(s_0)| = \sup_{|t| \leq T}|A(H+it)|$ on a segment satisfying \eqr{eq:comparison}. Since $\gamma<\gb$ on the positive edge, and by the triangle inequality,
\begin{equation}
 A_\edge(h_0)\leq\gb^d A_\edge(H),
 \qquad Q(h_0)\leq
 \frac{\gb^d A_\edge(H)}{h_0+1}+Q_\bulk(h_0).
 \label{v3:main-edge-compare}
\end{equation}
Inserting this into \eqr{bweak}, and using \eqr{edgepos}, \eqref{eq:comparison} and \eqref{abel}, we have (recalling $\lb = \l_1 \gb$)
\begin{equation}
 \begin{aligned}
 \lambda_1^d \lesssim c^p\left(\gb^d+\sqrt{c}\lambda_1^{3/2}\gb^{2-h_0}\right)
 \lesssim c^p\gb^d+\lambda_1^{p+2}
  \end{aligned}
 \label{eq:preliminary}
\end{equation}
where we have used $h_0\geq p + 5/2$. This implies, using $d>p+2$, 
\e{}{\lambda_1\lesssim c^{p/(2d)}\sqrt{\lb}}
Because $\lb$ has no upper bound, this preliminary estimate need not itself bound $\l_1$ non-trivially; instead its role is to sharpen the estimate on the bulk maximum in \eqr{abel}. 

In particular, substitution in \eqref{abel} for $h = H$ and using $\lb \gtrsim c$ gives
\begin{equation}
 Q_{\bulk}(H)\lesssim c^{-{(d-p-2)(d+p)/2d}}
 \label{eq:small}
\end{equation}
The right-hand side is $o(1)$, since $d>p+2$. Thus, for parameters in the prescribed range $d>p+2$, 
\e{bulko1}{Q_{\bulk}(H) = o(1)}
In other words, in a right half-plane $\Re(s)>H$, the bulk contribution $Q_\bulk(H)$ always vanishes as $c\rar\i$.

\eqr{bulko1} is a crucial observation. To see its effect, let us now take the point $s=H+it$ to be the boundary-maximizing point {\it of the full function $|A(s)|/|s+1|$} on $\Re(s)=H$, as defined in \eqr{bmax}; that is, we now sit at a new point on the same vertical line $\Re(s)=H$, at $t = t_*$. By definition of $Q(H)$ as a maximum, its value is at least the value of $|A(s)|/|s+1|$ at $t=0$. The reverse triangle inequality at zero, and positivity plus the triangle inequality at $t_*$, give
\begin{equation}
 \frac{A_\edge(H)}{H+1}-Q_\bulk(H)
 \leq\frac{|A(H+it_*)|}{|H+1+it_*|}
 \leq\frac{A_\edge(H)}{\sqrt{(H+1)^2+t_*^2}}+Q_\bulk(H)
 \label{v3:main-sandwich}
\end{equation}
where $A_\edge(H) \geq 1$. Therefore, $Q_\bulk(H) = o(1)$ implies $t_* = o(1)$. In other words, the maximum lives on the real line,
\e{}{t_* \rar 0\,,}
as $c\rar\i$. Therefore, plugging into \eqr{l1t} implies 
\e{}{\l_1 = O(1)\,.}

Finally, removing the assumption on $\lb$ implies $\l_1 = o(c)$. We argue by contradiction. If $\lambda_1\neq o(c)$, there is a subsequence with $\lambda_1\geq\varepsilon c$ for some fixed $\varepsilon>0$. Then $\lb>\lambda_1\geq\varepsilon c$, so the previous result applies on that subsequence and gives $\lambda_1=O(1)$, a contradiction. Thus,
\e{l1c}{\l_1 = o(c)\,.}
This concludes the proof. $\square$

\begin{figure}[t]
    \centering{\includegraphics[scale=.9]{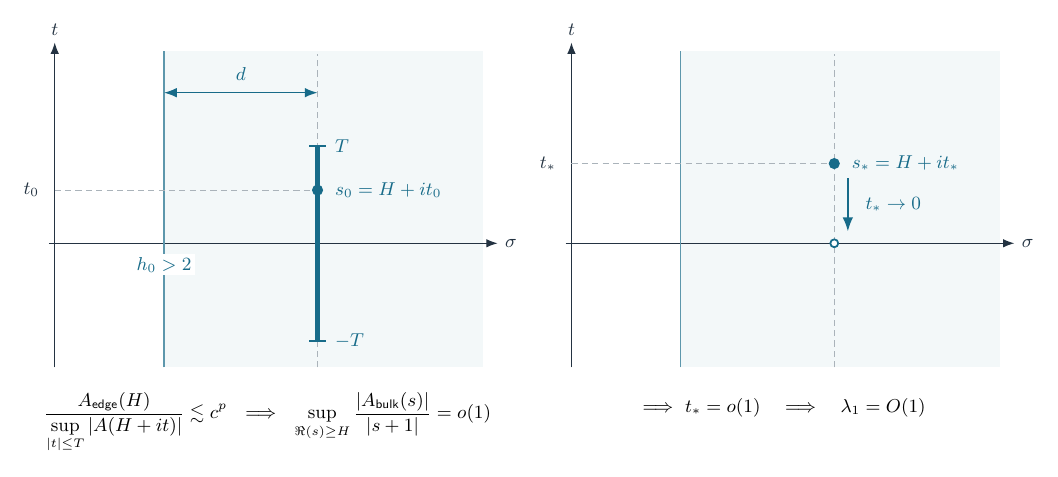}}
    \caption{A schematic visual aid to the two main steps of Section \ref{s421}'s proof. First, at the comparison point $s_0$ which maximizes $|A(H+it)|$ on a segment, one derives a preliminary bound on $\l_1$, which bounds the bulk contribution. This is then used to show, upon picking the point $s_*$, that the maximum of $|A(s)|/|s+1|$ on the half-plane $\Re(s)\geq H$ must live at $t_* \rar 0$  at large $c$. By \eqr{bstrong}, this yields a bound on $\l_1$.}
    \label{}
\end{figure}

\sssec{Comments}

\paragraph{Zero sparsity:} It is important to stress that nowhere did we make any explicit assumption about extra structure of the {\it zeros}. In fact, the preceding argument also controls the zeros of the completed function $\Xi(s)$, implying zero-free discs, and affirming zero sparsity \eqr{zscond}. To see this, we note that the fact that $t_* \rar 0$ as $c\rar\i$ implies that the zero sum in \eqr{rexibd} is uniformly upper-bounded: in particular,\foot{Assuming $\l_1>0$ uniformly in $c$, which is of course the range relevant for gaps.}
\e{}{0\leq \sum_\rho\frac{|H-\Re(\rho)|}{|s_*-\rho|^2} \leq {1\o 2r_H}}
where $s_* \rar H$ as $c\rar\i$, for a fixed positive constant $r_H$. Each term is non-negative, and is thus itself upper-bounded by $1/2r_H$, so for every zero,
\e{}{\qquad\qquad\qquad\quad |\rho-s_*|^2 \geq {2 r_H(\Re(\rho)-H)} \qquad (\Re(\rho) > H)}
Viewed as a constraint on $\rho$, this defines a uniform zero-free open disc centered at $H+r_H+it_*$, of radius $r_H$. Since $t_*\rar 0$, any strictly smaller disc centered at $H+r_H$ is eventually zero-free. This holds for all admissible $H$ in the proof. Similar algebra shows that \eqr{zscond} is satisfied. 

\paragraph{Numerical bound:} We can extract a numerical upper bound on $\l_1$ from the above proof. For simplicity, let us take $p=0$. Once the proof is complete, the final maximizing line can be moved left, because \eqr{abel} improves to $Q_{\bulk}(h) =O(c^{5/2-h})$ (by plugging in $\l_1 = O(1)$). Thus, $Q_{\bulk}(h)\to0$ at every fixed $h>5/2$, and the ``sandwich'' \eqr{v3:main-sandwich} gives $t_*(h)\to0$ on each such line. Consequently, we have a numerical upper bound on $\l_1$ from \eqr{bstrong}:
\es{}{\limsup_{c\to\infty}\lambda_1 &\leq \inf_{h>5/2} \exp\({F(h)+\frac1{h+1}}\)}
where we recall the definition of $F(s)$ in \eqr{Fdef}. One can check that the exponent increases monotonically, so,
\e{}{\limsup_{c\to\infty}\lambda_1 \leq \exp\(F\({5\o2}\)+{2\o 7}\) = \frac{\exp\({2\frac{\zeta '(5)}{\zeta (5)}+\frac{527}{105}-2 \gamma_E}\)}{8\pi^2} \simeq 0.57155}
where $\g_E$ is Euler's constant. This is irrelevant for the $\lb$-agnostic bound \eqr{l1c}, but bounds $\l_1$ by an explicit number along any putative sequences for which exceptional gaps to $\sim c$ are realized. This is just a partial optimization of the bound on $\l_1$; we do not expect the precise number above to hold deep significance. 

\sssec{A simpler logarithmic bound}\label{s423}
The intermediate results in the proof above easily imply a bound under weaker assumptions, by using the logarithmic bound \eqr{logbound}. In particular, let us {\it drop} the requirement $\lb\gtrsim c$, and only demand a fixed $H>2$ rather than \eqr{hrange}. We again choose $s_0=H+it_0$ to attain the supremum in \eqr{eq:comparison}. By the triangle inequality, $A_\edge(H)\ge1$, the relation \eqr{abel} and $\gamma_\sfb\ge1$, it is straightforward to see that
\begin{equation}
\begin{aligned}
\frac{Q(H)}{|A(s_0)|}
\lesssim c^p
\left(1+\sqrt{c}\,\lambda_1^{3/2}\right).
\end{aligned}
\end{equation}
Substituting into \eqr{logbound} gives, for fixed $H,T$,
\begin{equation}
\begin{aligned}
\lambda_1 &\lesssim \(\log c+\log(2+\lambda_1)\)^2\,.
\end{aligned}
\end{equation}
The last logarithm can be absorbed into the left-hand side, yielding $\lambda_1=O(\log^2 c)$, with implied constants independent of $c$. Having only demanded \eqr{Sass} for $\lb \gtrsim c$, the bound above should again be regarded as a bound $\l_1 = o(c)$. We will next use this same logarithmic bound strategy to derive $\l_1 = o(c)$ in the general case, in which the bulk reference density is nonzero. 

\sssec{General bulk density}\label{s424}

We now allow $\bar A_\bulk(s)\neq 0$, and bound $\l_1$ using the logarithmic bound \eqr{logbound}.\foot{We present the proof in purely analytic terms to emphasize its rigor, deferring its physical explanation to Section \ref{s5}. However, as a preview, it will be useful to think in terms of the dichotomy between the ``enigmatic'' region $\l \lesssim c/12$, in which non-universal large deviations from Cardy are possible, and the ``asymptotic'' region $\l \gtrsim c/12$ in which Cardy universality holds \c{HKS}.} Our goal is thus to constrain the size of the ratio $Q(h)/|A(s_0)|$ for points $s_0 = h+it$ with $A(s_0)\neq 0$. 

We initially take $h=H>\max\{1,a+\half\}$, and again choose $s_0=H+it_0$ to attain the supremum in \eqr{eq:comparison}. Define the half-plane maxima of the background and fluctuation parts of $A_\bulk(s)$ defined in \eqr{Abulkdeff}:
\begin{equation}
\begin{aligned}
Q_{\overline{\bulk}\,}(H)&\coloneq\sup_{\Re(s)\ge H}\frac{|\barAbulk(s)|}{|s+1|},\quad
\Qdelta(H)&\coloneq\sup_{\Re(s)\ge H}\frac{|\dAbulk(s)|}{|s+1|}\,,
\end{aligned}
 \label{eq:norms}
\end{equation}
As before, edge positivity implies 
\e{}{Q_\edge(H) = {A_\edge(H)\o H+1}\,.}
By the triangle inequality,
\e{Qtri}{{Q(H)\o |A(s_0)|} \leq {1\o |A(s_0)|} \(Q_\edge(H) + Q_{\overline{\bulk}\,}(H) + \Qdelta(H)\)}
The first and last terms are controlled. It remains to control the background contribution relative to $|A(s_0)|$, without bounding its absolute size. 

To do so we start by Abel summation. Let $\bar\nu$ be a locally finite signed measure on $[\lb,\infty)$, with cumulative count $\bar S(\l)=\bar\nu([\lb,\l))$, for each fixed $c$ (no lower or upper bound on $\lb$ is imposed). Integrating by parts yields the Abel representation,
\begin{equation}
 \bar A_{\bulk}(s)
 =\(s-\tfrac12\)\lambda_1^{s-1/2}
 \int_{\lb}^\infty \bar S(\l)\l^{-s-1/2}\,d\l
 \label{abelbar}
\end{equation}
Its validity and the convergence of the integral are given in Appendix \ref{appb}. Noting the elementary inequalities $|s-\half|\leq |s+1|$ and $H+1 \leq |s+1|$, the half-plane maximum obeys
\begin{equation}
 \Qbar(H)\leq \l_1^{H-1/2}\int_{\lb}^{\infty}|\bar S(\l)| \l^{-H-1/2} d\l
 \label{eq:ref-envelope}
\end{equation}
Now, suppose that $\bar S(\l)$ were sign-definite. In that case, the integrand in \eqref{abelbar} at the {\it real} point $s=H$ would have that sign: consequently,
\begin{equation}\label{Abarbulksign}
 |\barAbulk(H)|=\(H-\tfrac12\)\l_1^{H-1/2}
 \int_{\lb}^{\infty}|\bar S(\l)|\l^{-H-1/2} d\l\,.
\end{equation}
This implies $\Qbar(H) \leq |\barAbulk(H)|/(H-\half)$. For our purposes, literal sign-definiteness is stronger than necessary: we can allow a more forgiving condition on $\bar S(\l)$ which permits polynomial loss envelopes in $c$,
\e{Qboundpoly}{\Qbar(H)\le\frac{U_c}{H-\frac12}|\barAbulk(H)| + V_c\,,\quad 1\lesssim U_c \lesssim c^u\,,\quad |V_c| \lesssim c^v\,.}
This would follow from sign-definiteness of $\bar S(\l)$ (corresponding to $U_c=1, V_c=0$), but also allows $\bar S(\l)$ to have opposite-sign excursions of polynomial size that would augment the right-hand side of \eqr{Abarbulksign} with $V_c$. From a purely analytic point of view, this property of $\bar S(\l)$ is a simplifying technical assumption. However, it holds for large $c$ scalar primary densities of $Z_p(\t)$ and $\tilde Z_p(\t)$, as we explain in Appendix \ref{appb}. Thus, we impose \eqr{Qboundpoly} as a sufficient analytic condition on the cumulative difference $\bar S(\l)$ of the reference densities.

From here, repeated use of triangle inequalities and the relation $A_\bulk(s) = A(s) - A_\edge(s)$ imply
\e{Qratio}{{Q(H)\o |A(s_0)|} \leq {U_c\o H-\tfrac12} + \({U_c\o H-\tfrac12} + {1\o H+1}\){A_\edge(H)\o |A(s_0)|} + \(1+U_c{H+1\o H-\tfrac12}\) {Q_\d(H)\o |A(s_0)|} + {V_c\o |A(s_0)|}}
The contribution from $\bar A_\bulk(s)$ has been transmuted into contributions to the edge and fluctuation terms. Recalling that $s_0$ is chosen to be a point at which \eqr{eq:comparison} holds, and using \eqr{Sass} and $A_\edge(H)\geq 1$, the same calculation leading to \eqr{abel} gives\foot{The next section is devoted to explaining why the pointwise fluctuation bound \eqr{Sass} holds in the CFTs of interest, as a spectral rigidity property, but we also note that \eqr{Sass} is sufficient, not necessary, for deriving \eqr{l1boundoc}. A direct polynomial estimate on the residual maximum $\Qdelta(H)$ also suffices; see Section \ref{s531} for a statistical version of this.}
\e{Qratio2}{{Q(H)\o |A(s_0)|} \lesssim c^{\hat p}(1+c^b\l_1^a)}
where $\hat p=p+\max\{u,v\}$ is fixed. Plugging into \eqr{logbound}, the $\log\l_1$ absorbs, yielding $\l_1 = O(\log^2 c)$. As before, taking our assumption \eqr{Sass} to hold for $\lb\gtrsim c$ yields the bound
\e{l1boundoc}{\l_1 = o(c)\,.}
This concludes the proof in the general case. $\square$

\sec{Spectral statistics and log-rigidity}\label{s5}

A central ingredient of the proof is the counting fluctuation bound \eqr{Sass}. Here we explain its origin from a spectral statistical viewpoint: in short, a sufficient condition is a {\it logarithmic bound on counting fluctuations}, inherited from random matrix statistics. Actually, a rather weaker form of this is also sufficient for \eqr{Sass}, as we show below. To be clear, the preceding proof is fully deterministic; the purpose of this section is to derive \eqr{Sass} from statistical properties of the CFT spectral universality class of interest. We conclude with a remark on the regime of validity of our fluctuation bound. 

\ssec{Warmup motivation: RMT}\label{s51}

The counting hypothesis is motivated by familiar properties of random matrix statistics. Consider a local statistical sampling of a deterministic individual spectrum, with mean density $\rho_0(\lambda)$. Introduce the unfolded coordinate and density fluctuation over a spectral interval $[\l_*,\l]$, 
\e{}{u(\lambda)=\int_{\lambda_*}^{\lambda}\rho_0(E)\,dE,
\qquad\delta\mathsf{\rho}(u)
=\frac{\rho(\lambda(u))}{\rho_0(\lambda(u))}-1.}
The mean unfolded density is one. Assuming stationarity within the intervals considered, define the connected correlator and its Fourier transform by
\e{}{
\mathsf C_\rho(u-u')
\coloneq\mathbb E\!\left[\delta\mathsf\rho(u)\delta\mathsf\rho(u')\right],
\qquad
\mathsf C_\rho(r)
=\int_{\mathbb R}\frac{dk}{2\pi}\,e^{ikr}m_\rho(k)}
where $\mathbb{E}$ is the expectation value induced by the local sampling within the stationary window (or, in actual random matrix theory, by draws of the matrix). Here $m_\rho$ is the structure factor, otherwise known in high-energy parlance as the microcanonical spectral form factor. For an interval of unfolded length $\ell$, its number variance is
\e{}{
\begin{aligned}
\Sigma^2(\ell)
&=\int_0^\ell du\int_0^\ell du'\,\mathsf C_\rho(u-u')=\frac1\pi\int_{\mathbb R}dk\,
\frac{1-\cos(k\ell)}{k^2}\,m_\rho(k).
\end{aligned}}
Whereas a system with Poisson statistics obeys $m _\rho(k) = 1$ and $\Sigma^2(\ell) = \ell$, random matrix statistics give
\e{}{\Sigma^2(\ell) \sim \log\ell\,.}
For the GUE, for example, 
\e{}{m_\rho(k)=\min\!\left\{\dfrac{|k|}{2\pi},1\right\},\qquad \Sigma^2(\ell)=\dfrac{1}{\pi^2}\log\ell+O(1).}
The linear ramp produces logarithmic growth of the number variance. Other symmetry classes retain this logarithmic growth as $\ell\gg1$, with a symmetry-dependent coefficient. The typical number fluctuation is thus $O(\sqrt{\log\ell})$. 

Now, the statistical {\it maximum counting discrepancy} over an interval is larger, in general, due to rare events: this is the purview of extreme value theory and large deviations. However, by a result of \c{BLZ} for random matrices in various ensembles, including the standard GUE, GOE and GSE cases, the extreme values are merely logarithmic, i.e. linear in $\Sigma^2(\ell)$.\foot{The behavior \eqr{rmtopt} is actually just the leading order of a more refined property of random matrix extreme value statistics, governed by the Fyodorov-Hiary-Keating conjecture \c{FK, FHK} and its extensions \c{ABB15,PZ16,CMN,FLD,PZ22,BLZ}. For an introduction to these ideas in an AdS/CFT context, and further references, see \c{2607}.} Defining the maximal deviation as
\e{}{M_\ell \coloneq \sup_{0\le u\le\ell}\bigl|\cN([0,u))-u\bigr|}
the result is
\e{rmtopt}{{M_\ell\o \log\ell} \xlongrightarrow{\mathbb P}
\frac{1}{\pi}\sqrt{\frac{2}{\beta}}}
as $\ell\rar\i$. Casually paraphrasing, in systems with random matrix statistics on growing unfolded intervals, large-scale counting fluctuations around the average are bounded in probability by a logarithm. The result \eqr{rmtopt} was also established in \c{Holcomb:2018SineMaximum} for the Sine$_\b$ process, to which the unfolded local bulk statistics of $\b$-ensembles converge in the large $N$ limit.

\subsection{Log-rigidity of CFT spectra}
Our goal here is to state a sufficient condition on the class of partition functions we wish to consider. We then return to the differences $\cZ(\t)$, and the series $A(s)$, afterwards.

Given a unitary primary partition function $Z_p(\t)$, let $\cN(I)$ count its scalar levels in a $\l$-interval $I$, including multiplicity. Fix a physical width $L>0$. For some prescribed value $\lb>0$, partition $\l \geq \lb$ into consecutive cells $I_j=[u_j,v_j)$ of width $L\coloneq v_j-u_j$ for $j>0$, allowing a shorter first cell (which we denote $I_0$). We call this a ``covering'' of the bulk spectrum. Defining subintervals $I_j(\l) \coloneq [u_j,\l) \subseteq I_j$, the quantity 
\e{}{M_{j}\coloneq\sup_{u_j \leq \l \leq v_j}\Bigg|\cN(I_j(\l))-\int_{I_j(\l)}\rho_{0}(\l')\,d\l'\Bigg|}
measures the largest deviation over all subintervals of $I_j$ starting at $u_j$ from the unfolded average computed with reference density $\rho_0(\l)$. The unfolded length of the full interval $I_j$ is
\e{}{\ell_j\coloneq\int_{u_j}^{v_j}\rho_{0}(\l)\,d\l\,.}
The main point can now be made. A sufficient condition on the theories under consideration is what we call a {\it log-rigidity condition} on the scalar spectrum: there exists a covering of the bulk range $\l\geq \lb \sim c$ for which, on each interval $I_j$, 
\e{logcft}{M_j \leq K \log(\ell_j+2)}
with $K$ independent of $j$. 

This is the enforcement of the optimal rigidity result \eqr{rmtopt} familiar from random matrix theory, for each interval $I_j$ in the bulk spectrum. It is not guaranteed purely from local random matrix statistics in a single interval, as we ask for a good covering of the bulk over all intervals. We explain how to quantify this assumption in Appendix \ref{appa}. But neither is it a necessary condition for the physics we will later want: in particular, we have made some choices above which may be relaxed. (For instance, there is no need to choose every interval to have literally the same $O(1)$ length $L$, so one could let the lengths vary as well in seeking a good covering.) \foot{Note that the condition \eqr{logcft} is not exclusive to random matrix statistical universality classes: since we impose only an upper bound, more rigid spectra are allowed as well. Spectra that are more rigid than logarithmic (on local scales) are known as Class I hyperuniform spectra. A hyperuniform spectrum is defined as having a number variance within a region $\mathcal{B}$ that enjoys sub-volume scaling in the limit Vol$(\mathcal{B})\rar\i$. We are not aware of the appearance of hyperuniformity in a CFT context, but we point out that hyperuniformity comprises a large and active subfield of statistical physics \c{Torquato:2018Hyperuniform}. It is interesting to ask whether any CFTs have high-energy spectra which exhibit hyperuniform scaling of Class I or III.} 

By summing over cells, one can compute the upper envelope of the cumulative deviation over a given range $[\lb,\l)$,
\e{}{\sum_{j=0}^{j_*(\l)} M_j \leq K \sum_{j=0}^{j_*(\l)}\log(\ell_j+2)}
where $j_*(\l)$ is defined by the condition $\l\in I_{j_*(\l)}$ (i.e. it is the index of the interval containing $\l$).

The reference density $\rho_0(\l)$ can be fixed by a prescription for modular completion of the light spectrum of $Z_p(\t)$, possibly augmented by theory-dependent contributions from enigmatic states. It is important to stress that we will not need to assume a specific form for this density, as we see very shortly. However, modular invariance implies that we may think of $\rho_0(\l)$ as the scalar Cardy formula plus other terms. Let us record the Cardy formula here: from the $S$-modular transform of the vacuum module, as a density in $\l$, 
\e{Cardydef}{\rho_{\rm Cardy}^{(0)}(\l) \coloneq {4\o P^2}\sinh^2(2\pi b P)\sinh^2(2\pi b^{-1}P) \approx {1\o 2\l} e^{2\pi \sqrt{{c-1\o 3}\l}}}
where $\l = 2h -{c-1\o 12}= 2P^2$ with $P\in\RR$ and $c=1+6(b+b^{-1})^2$.

\ssec{From log-rigidity to the counting bound}\label{s53}

We now return to the setting of our bootstrap, in which we consider the difference $\cZ(\t) = Z_p(\t) - \widetilde Z_p(\t)$. In terms of the constituent spectra, the signed count of states $\nu(I)$ over intervals $I$ is
\e{}{\nu(I) = \cN(I) - \widetilde \cN(I)\,.}
Recalling earlier definitions, this admits a decomposition into background and fluctuation parts, defined with respective reference densities introduced above:
\e{}{\nu(I) = \bar\nu(I) + \d\nu(I)\,,\quad \text{where}\quad \bar\nu(I) \coloneq \int_I \(\rho_0(\l) - \widetilde\rho_0(\l)\) d\l}
and likewise for the cumulative signed count of bulk states, $S(\l) = \nu([\lb,\l)) = \bar S(\l) + \d S(\l)$.

\sssec{Fluctuations}\label{s531}

Summing the centered counting discrepancies along their respective coverings, and then cancelling the reference contributions, bounds the cumulative signed count $\d S(\l) = \d\nu([\lb,\l))$: by the triangle inequality, 
\e{goodcover}{|\d S(\l)| \leq \sum_{j=0}^{j_*(\l)}M_j + \sum_{\tilde j =0}^{\tilde j_*(\l)}\widetilde M_{\tilde j}}
where $M_j$ and $\widetilde M_{\tilde j}$ are suprema over respective coverings $\{I_j\}$ and $\{\tilde I_{\tilde j}\}$. In particular, when $M_j$ and $\widetilde M_{\tilde j}$ obey the log-rigidity property \eqr{logcft}, $|\d S(\l)|$ is effectively logarithmic, as it is bounded by sums of logarithms of the respective unfolded cell lengths. 

To finally pass to \eqr{Sass} -- and to explain why a vast generalization of \eqr{logcft} is also acceptable -- we note that any reference density growing exponentially in powers of $c$ and $\l$ yields \eqr{Sass}. We have 
\e{eqeqeq}{\log(\ell_j+2) \leq \log\(L\sup_{\l\in I_j}\rho_0(\l)+2\)}
Plugging into \eqr{goodcover} yields \eqr{Sass}, for any reference density with upper envelope
\e{rhoenv}{\log\rho_0(\l) \leq C c^b \l^{a-1}}
for $\l\geq \lb$, where $C$ is a coefficient independent of $c$ and $\l$. 

This makes it immediately clear that the condition 
\e{logrigidqr}{M_j \lesssim c^q\log^{r}(\ell_j+2)}
for $q,r\in\R_+$ would also lead to \eqr{Sass}: poly-logarithmic rigidity, or even power law growth with $c$ of the counting fluctuations, would preserve it. This significantly expands the class of partition functions to which our spectral bound \eqr{l1boundoc} applies (even if \eqr{logrigidqr} lacks statistical motivation for $q>0$ or $r>1$). 

\sssec*{A statistical alternative to log-rigidity}

As an aside, instead of extremal log-rigidity of the underlying spectra, a weaker {\it statistical} input can directly control $Q_\delta(H)$, without establishing the deterministic counting bound \eqr{Sass}. Let $\EE$ and $\PP$ refer to a statistical model of the signed spectrum that is induced by sampling. Suppose that the fluctuation field $\d S(\l)$ satisfies the endpoint estimate
\begin{equation}
 \bigl(\EE\[|\delta S(\l)|^2\]\bigr)^{1/2}
 \le Cc^b \l^a,\qquad \l\ge\lb,
 \label{eq:endpoint-moment}
\end{equation}
with fixed $a,b\ge0$ and $C$ independent of $c$ and $\l$. This controls each endpoint in mean square, without controlling the maximum over endpoints. Logarithmic number variance, as in Section \ref{s51}, motivates this input. Summing these residuals over the intervals $\{I_j\}$ gives a polynomial envelope of the form \eqref{eq:endpoint-moment}. For fixed $H>\max\{1,a+\tfrac12\}$, Abel and Minkowski's inequality (i.e. the triangle inequality on $L^2$ norms) give
\begin{equation}
\begin{aligned}
 \bigl(\EE\[\Qdelta(H)^2\]\bigr)^{1/2}
 &\le \lambda_1^{H-1/2}\int_{\lb}^{\infty}
       \bigl(\EE\[|\delta S(\l)|^2\]\bigr)^{1/2}\l^{-H-1/2}\,d\l \leq C_H c^b\lambda_1^a
 \label{eq:statistical-abel}
\end{aligned}
\end{equation}
where we used $\lb\ge\lambda_1$, and absorbed trivial $H$-dependence into the constant $C_H$. This is precisely the scaling used in the proof. Note further that for every fixed $\eta>0$, Chebyshev's inequality then yields
\begin{equation}
 \PP\!\left(\Qdelta(H)>c^{b+\eta}\lambda_1^a\right)
 \le C_H'c^{-2\eta}.
 \label{eq:statistical-probability}
\end{equation}
The way to read the implication of this statistical constraint is that the bootstrap result of Section \ref{s4} holds with probability tending to one (whenever the remaining analytic assumptions hold). 

\sssec{Backgrounds}\label{s532}

The preceding discussion controls fluctuations around the respective reference densities, but does not require those densities to coincide. The robustness of our proof to the choice of reference densities is important, as we now explain. 

First, consider the case where the references {\it do} coincide, such that $\rho_0(\l) = \widetilde\rho_0(\l)$, and hence $\bar A_\bulk(s)=0$. This is the case if we fix a modular completion prescription for the reference densities of $Z_p(\t)$ and $\widetilde Z_p(\t)$: since their light spectra are identical, their references cancel exactly, and $\bar S(\l)=0$ over the whole bulk. The whole calculation is controlled by fluctuations. Therefore, in view of the preceding subsection, the bound on $\l_1$ established in Section \ref{s421} holds when both scalar spectra are log-rigid around the same density (for instance, $\rho_0(\l) = \widetilde\rho_0(\l) = \rho_{\rm Cardy}^{(0)}(\l)$). 

From this point of view, one reason we must consider $\bar A_\bulk(s)\neq 0$ (cf. Section \ref{s424}) is that non-universal deviations from Cardy are allowed in the ``enigmatic'' regime $\l-{1\o 12} \in[0,{c\o 12}]$ \c{HKS}. For a sparse light spectrum at large $c$, the leading entropy is universal above this range, and upper-bounded within it. In particular, for an appropriately averaged density of states $\rho(\l)$ summed over all spins (see \c{Mukhametzhanov:2019pzy} for details on the averaging procedure), 
\es{}{\log\rho(\l) &= 2\pi\sqrt{c\l\o 3} + o(c)\,, \quad \qquad \l\gtrsim {c \o 12}\\
\log\rho(\l) &\leq {\pi c\o 6} + 2\pi\l + o(c)\,,\qquad\,  \l \lesssim {c\o 12}\,,}
with $\l/c$ held fixed as $c\rar\i$. By unitarity, the same upper bound applies to the averaged scalar density, permitting entropies to exceed the scalar Cardy entropy \eqr{Cardydef} by order $c$. 

The log-rigidity of fluctuations, with density obeying \eqr{rhoenv}, comfortably accommodates {\it any} allowed enigmatic scaling (as does its statistical alternative \eqr{eq:endpoint-moment}). The work of Section \ref{s424} and Appendix \ref{appb}, therefore, was to establish that even when enigmatic states give rise to a large difference $\rho_0(\l) - \widetilde\rho_0(\l)$, the bound on $\l_1$ holds. In particular, the bound does not rely on restricting our attention to CFTs which are log-rigid around any {\it specific} reference density (such as Cardy). The reason is that the proof only needs control of the ratio appearing in \eqr{Qtri}, not the absolute size of $\bar S(\l)$, and the former is bounded by \eqr{Qboundpoly} and \eqr{Qratio2}. 

\ssec{A sharper bound from extended log-rigidity?}

Let us return to our proof of \eqr{l1boundoc} and make the following remark. The analytic argument did not require the bulk threshold to be extensive in $c$, i.e. to obey $\lb \gtrsim c$. Suppose instead that
\begin{equation}
 \lb\gtrsim c^\alpha
 \label{eq:extended-threshold}
\end{equation}
for fixed $\alpha>0$. If we retain our setup and assumptions for the edge and bulk, the auxiliary lines may be shifted by $\a$-dependent values such that the same proof again gives $\l_1=O(\log^2 c)$ (or, in the setup of Section \ref{s421}, $\l_1 = O(1)$). As before, this also gives a conclusion without prescribing the scaling of $\lb$. Suppose that, for every fixed $0<\alpha<1$, the preceding hypotheses hold uniformly along any subsequence on which $\lb\geq c^\alpha$. Given
$\varepsilon>0$, choose $0<\alpha<\min\{\varepsilon,1\}$. Where $\lb<c^\alpha$, the ordering $\l_1<\lb$ gives $\l_1/c^\varepsilon\to0$; on the complementary subsequence, the
fixed-$\alpha$ argument gives $\l_1=O(\log^2 c)$. Consequently, this would imply
\e{l1strong}{\l_1 = o(c^\vareps)~~\forall\,\vareps>0\,.}
This would still leave open the possibility of $\l_1$ growing more slowly than every positive power of $c$.

The physical question is, therefore, how far below energies of order $c$ the bulk fluctuation hypothesis \eqr{Sass} should be imposed. In particular, to the extent that the log-rigidity input is motivated by tightness of the CFT spectra to random matrix statistics, it is unclear whether its extension to $\a<1$ is physically reasonable over a wide class of CFTs. Some reason for caution is afforded by the AdS$_3$ bulk point of view. On the one hand, states with $\l \sim c^\a$ with $\a<1$ are dual to black holes that remain parametrically lighter than the Planck scale; on the other hand, they are not large black hole states. Explicitly, the outer horizon radius $r_+$ has scaling
\begin{equation}
 \frac{r_+}{L_{\mathrm{AdS}}}\sim c^{(\alpha-1)/2},
 \qquad
 \frac{r_+}{G}\sim S\sim c^{(1+\alpha)/2}\,.
 \label{eq:extended-btz-scales}
\end{equation}
For $0<\alpha<1$, these states are sub-AdS but are not yet truly quantum black holes, as they still have parametrically large entropy in Planck units. Still, statistical expectations in this regime are less clear. 

Altogether, the stronger result \eqr{l1strong} seems, in our view, potentially unjustified. We therefore view it as a strengthening for putative classes of CFT pairs with the corresponding extended regime of validity of log-rigid counting control.

\sec*{Acknowledgments}

We thank the organizers of Strings 2025, Abu Dhabi, where an early version of this work was announced. We are grateful to Anshul Adve, Nathan Benjamin, Scott Collier, Tom Hartman, Simeon Hellerman, Henry Maxfield, Dalimil Maz\'a\v{c}, Hirosi Ooguri, Sridip Pal, Boris Post, Edgar Shaghoulian, Steve Shenker and Douglas Stanford for helpful discussions. We acknowledge the use of ChatGPT for assistance in honing arguments, making figures and draft editing. We thank the IHES for hospitality during the course of this work. This work was made possible by Institut Pascal at Universit\'e Paris-Saclay with the support of the program ``Investissements d'avenir'' ANR-11-IDEX-0003-01. This work was performed in part at the Aspen Center for Physics, which is supported by National Science Foundation grant PHY-2210452. We would like to thank the Isaac Newton Institute for Mathematical Sciences, Cambridge, for support and hospitality during the programme ``Physics and automorphic L-functions'' where work on this paper was undertaken. This work was supported by EPSRC grant no EP/Z000580/1, and in part by ERC Starting Grant 853507.

\begin{appendix}

\sec{Good covering}\label{appa}

The existence of a good covering is a deterministic statement. One sufficient way to establish it is as follows. Fix a physical width $L>0$, independent of $c$. We choose $U$ uniformly in $[0,L)$, use $I_0(U)=[\lb,\lb+U)$, and continue with width-$L$ cells to generate the complete partition
\begin{equation}
 I_0(U)=[\lb,\lb+U),\qquad
 I_j(U)=[\lb+U+(j-1)L,\lb+U+jL),\quad j\ge1.
 \label{v3:sampling-cover}
\end{equation}
Write its endpoints as $u_j(U),v_j(U)$. Every choice of $U$ covers the same bulk, and leaves $A_\bulk$ unchanged. In this setup, a sufficient ``transfer'' assumption is the following. For a fixed $K$, let $E_{c,j}(K)$ be the event $M_j(U)>K\log(2+\ell_j(U))$ -- that is, the violation of log-rigidity (for that chosen value of $K$, at least). Suppose that, for sufficiently large $c$, 
\begin{equation}
 \mathbb{P}_U(E_{c,j}(K))\le \varepsilon_{c,j},
 \qquad
 \sum_{j\ge0}\varepsilon_{c,j}<1.
 \label{v3:sampling-transfer}
\end{equation}
Then a union bound gives
\begin{equation}
 \mathbb{P}_U\bigl(M_j(U)\le K\log(\ell_j(U)+2)~\forall\,j\bigr)>0.
 \label{v3:sampling-good-cover}
\end{equation}
The statement is that if the cell-wise violation probabilities are {\it summable}, then a good partition exists for every sufficiently large $c$, and we use that value of $U$ in our construction. (The sum in \eqref{v3:sampling-transfer} need not tend to zero: positive probability of success would suffice.) The above condition is a reasonable expectation when deviations from random matrix statistics would be expected to vanish at larger $j$ -- that is, higher energies. In general, this is the case.\foot{In \c{cmc}, the authors engineered an agentic algorithm for constructing unitary, modular-invariant functions, including those with gaps parametrically above $c/12$ in the large $c$ limit. The details are still being fully verified. One can confirm that for the cases treated in some detail in \c{cmc}, the spectra have exponentially large exact degeneracies, strongly violating the logarithmic counting condition \eqr{logcft}. Such fine-tuning appears to extend to the general construction. We thank Chi-Ming Chang for a discussion.}

This is just an example: the sampling scheme is not part of the intrinsic hypothesis. In particular, we have chosen identical-length intervals, but allowing $L_-\le v_j-u_j\le L_+$ with fixed $0<L_-\le L_+<\infty$ preserves the cumulative counting estimate and its exponents. 

\sec{Properties of background densities}\label{appb}

This appendix is devoted to establishing some properties of the background term $\bar A_\bulk$ used in the main text. Part \hyperref[appb1]{B.1} justifies the inequality \eqr{Qboundpoly} in the presence of enigmatic scalar states over a range $[\lb,\kappa c)$, for any fixed $\kappa$. (The natural value of $\k$ suggested by \c{HKS} is $\k=1/12$, but we keep $\k$ arbitrary.) Part \hyperref[appb2]{B.2} then extends this to the general case, and establishes convergence properties of $\bar A_\bulk$.

\ssec*{B.1~~~Enigmatic states and positivity}\label{appb1}

The setup for this subsection is as follows. We split the range $\l\geq \lb$ into an {\it enigmatic} region $\l\in [\lb,\l_\sfa]$, and an {\it asymptotic} region $\l>\l_\sfa$, with
\e{}{\l_\sfa\coloneq\max\{\lb,\kappa c\}\,,\quad \kappa>0~ \text{fixed}\,. }
We may depict this like so:
\begin{figure}[h]
    \centering{\includegraphics[scale=.8]{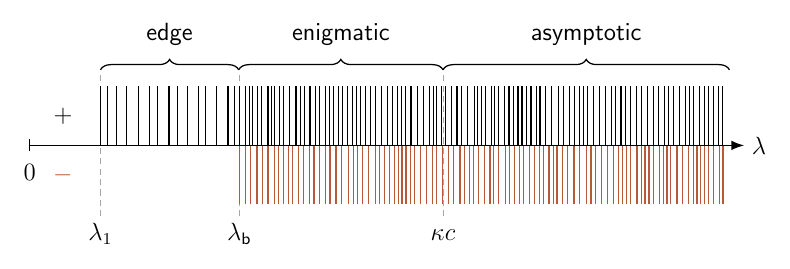}}
    \label{enigfig}
\end{figure}

\noindent In the asymptotic region, we take the signed counting discrepancy $\bar S(\l)$ accumulated from $\las$ to obey the polynomial fluctuation bound (a property we will relax later), whereas in the enigmatic region, we allow arbitrary, non-universal deviations from Cardy compatible with sparseness \c{HKS}. When $\lb > \kappa c$, the entire bulk is asymptotic. We emphasize that in the enigmatic region, the {\it absolute} size of $\bar S(\l)$ is unconstrained: in particular, it is allowed to grow exponentially with $c$, the maximal behavior allowed by \c{HKS}.

Let $\rho_0(\l)$ and ${\widetilde\rho}_0(\l)$ be the reference densities of the scalar primary spectra of $Z_p(\t)$ and $\tilde Z_p(\t)$, respectively. As explained above, we take their difference to be supported on $\l\in[\lb,\las]$, neglecting the polynomial differences at $\l>\las$ that are irrelevant here. Recalling our earlier definitions,
\begin{equation}
 \barAbulk(s)=\int_{\lb}^{\las}
 \left(\frac{\lambda_1}{\lambda}\right)^{s-\frac12}
 [\rho_0(\lambda)-{\widetilde\rho}_0(\lambda)]\,d\lambda,
 \qquad
 \Qbar(H)=\sup_{\Re s\ge H}\frac{|\barAbulk(s)|}{|s+1|}.
 \label{eq:definitions}
\end{equation}
For a fixed admissible $H$, the comparison used in the proof is \eqr{Qboundpoly}. Taking $V_c=0$ there for simplicity of presentation, the following density condition would be sufficient for \eqr{Qboundpoly}:
\begin{equation}
 \int_{\lb}^{\las}\left(\frac{\lambda_1}{\lambda}\right)^{H-\frac12}
 |\rho_0(\lambda)-{\widetilde\rho}_0(\lambda)|\,d\lambda
 \le U_c|\barAbulk(H)|\,.
 \label{branches}
\end{equation}
Indeed, for $\Re s\ge H$, the weight has modulus at most its value at $H$, and $|s+1|\ge H+1$, so $ \Qbar(H)$ is upper-bounded by the above integral. 

The essential point is that expanding the densities as sums of exponential contributions dressed by perturbative large $c$ expansions, the above integral will be dominated by the neighborhood of the largest exponential in the range $[\lb,\las]$. Define rescaled frequencies
\begin{equation}
 \l = y c\,,\quad \lb=\k_\sfb c,\qquad  0<\k_\sfb<\k<\infty,
 \label{eq:scales}
\end{equation}
so $\las = c\max\{\k_\sfb,\kappa\}$. Expanding the density difference, 
\begin{equation}
 {\rho}_0(yc)-{\widetilde\rho}_0(yc)
 =\sum_{j\in\Z_+}\epsilon_j c^{p_j}a_j(c,y)e^{c s_j(y)}\,,
 \qquad \epsilon_j\in\{+1,-1\}\,.
 \label{branches2}
\end{equation}
The entropy profiles $s_j(y)$ are assumed continuous on $[\k_\sfb,\k]$. As proven in \c{HKS}, $s_j(y)$ is independent of $c$. The powers $p_j$ are also $c$-independent. In gravitational terms, this is a sum over bulk saddles dressed by perturbative $1/c$ expansions. The $j$'th ``saddle'' has a leading perturbative expansion
\begin{equation}
 a_j(c,y)=a_{j,0}(y)+o(1),
 \label{eq:prefactors}
\end{equation}
for some coefficient $0<a_{j,0}(y)<\i$, and an asymptotic expansion to all orders in $1/c$ beyond. If we now define the largest entropies of each sign over the interval,
\begin{equation}
  s_\pm^*\coloneq\sup_{j:{\epsilon_j=\pm1}}\ \max_{y\in[\k_\sfb,\k]}s_j(y),
 \label{eq:entropy-maxima}
\end{equation}
then a sufficient condition for \eqr{branches} is, up to exchanging the two signs,
\begin{equation}
 {s_+^*\ge s_-^*+\delta\,,\qquad \delta>0~ \text{fixed}.}
 \label{eq:global-gap}
\end{equation}
The proof of \eqr{branches} from \eqr{eq:global-gap} is straightforward: one can show that $U_c = 1 + O(c^P e^{-c\delta/2})$ for some fixed $P>0$ (which absorbs polynomial factors in $c$), with implied constant independent of $c$, by defining the Mellin integrals
\begin{equation}
 I_\pm(H)\coloneq\sum_{\epsilon_j=\pm1}\int_{\k_\sfb}^{\k}
 \left(\frac{\lambda_1}{c y}\right)^{H-\frac12}
 c^{p_j+1}a_j(c,y)e^{c s_j(y)}\,dy.
 \label{eq:sign-integrals}
\end{equation}
in terms of which $\bar A_{\bulk}(H)=I_+(H)-I_-(H)$.\foot{If the leading entropies are instead ``tied'' such that $\d=0$, and the integrals $I_\pm(H)$ of their respective perturbative expansions $c^{p_j}a_j(c,y)$ differ at some finite order, the argument proceeds identically at the cost of a polynomial-in-$c$ factor in $U_c$.} 

Note that no assumption of monotonicity on the entropies $s_j$ is required, nor is it required that one branch dominate the entire interval $[\k_b,\k]$. Indeed, let us demonstrate this with an illustrative bottom-up example in which enigmatic saddles exchange dominance, and the cumulative count $\bar S$ even violates sign-definiteness, but \eqr{branches} is obeyed as claimed. 

\paragraph{Example:} Consider the two entropy profiles
\begin{equation}
s_1(y)=\frac{\pi}{6}\sqrt{1+12y}, \qquad s_2(y)=\frac{\pi\eta}{6}+2\pi y, \qquad \sqrt2-1<\eta<1\,.
\end{equation}
The first has a square-root dependence, characteristic of enigmatic black hole entropies (see e.g. Section 4.3 of \c{HKS}), while the second has Hagedorn growth. Both are increasing and satisfy the HKS entropy bound $s_i(y)\le\pi/6+2\pi y$ on $0\le y\le1/12$. We have
\begin{equation}
\begin{aligned}
s_1(0)-s_2(0)
&=\frac{\pi}{6}(1-\eta)>0,\\
\delta \coloneq s_2(1/12)-s_1(1/12)
&=\frac{\pi}{6}(1+\eta-\sqrt2)>0,\\
s_2'(y)-s_1'(y)&=2\pi-\frac{\pi}{\sqrt{1+12y}}>0.
\end{aligned}
\end{equation}
There is therefore one crossing: the black hole branch wins at the lower endpoint, and the Hagedorn branch wins at the upper endpoint, with a crossing at $y_* = (1+\sqrt{5-4\eta}-2\eta)/24$. 

We take our enigmatic interval to be 
\e{}{y\in[\k_\sfb,1/12]\,, \quad 0<\k_\sfb < y_*~\text{fixed},}
where the lower edge lies below the entropy crossing, and identify the densities $\rho_0$ and $\widetilde\rho_0$ with these entropy profiles:
\begin{equation}
\rho_0(yc)-\widetilde\rho_0(yc) = P(c,y)\(e^{cs_1(y)} - e^{c s_2(y)}\)\,.
\end{equation}
A smooth prefactor $P$ admits a uniform $1/c$ expansion (with leading term strictly bounded away from zero). This difference is positive below the crossing and negative above it; nevertheless, since both entropies are maximized at the upper endpoint $y=1/12$, the respective contributions to $\bar A_\bulk(s)$ are each exponentially dominated at the upper endpoint. Consequently, $s_2$ wins: computing their respective integrals \eqr{eq:sign-integrals}, the finite entropy gap $\delta>0$ implies 
\begin{equation}
\frac{I_+(H)}{I_-(H)}=O(e^{-c\delta}).
\end{equation}
$\bar A_{\bulk}(H)$ is exponentially dominated by the Hagedorn contribution, and 
\begin{equation}
Q_{\overline{\bulk}}(H) \le\frac{1+O(e^{-c\delta})}{H+1} |\bar A_{\bulk}(H)|.
\end{equation}
Thus, despite the fact that the cumulative reference count $\bar S(\l)$ also changes sign -- it is initially positive, but its value at $\l=c/12$ is negative for sufficiently large $c$ -- \eqr{branches} holds. 

\ssec*{B.2~~~Convergence and general reference densities}\label{appb2}
The previous section restricted the super-polynomial fluctuations of the reference densities $\rho_0$ and $\widetilde\rho_0$, and hence of the cumulative difference count $\bar S(\l)$, to lie in the enigmatic range. Here we show that modularity for smooth asymptotic expansions forbids super-polynomial fluctuations from extending to $\l\rar\i$ at fixed $c$. Throughout this subsection, $O(\cdot)$ indicates that implicit constants may depend arbitrarily on $c$.

We are interested in the behavior of $\bar S(\l)$ as $\l\rar\i$. As recalled in Section \ref{s2}, $\cZ(\tau)$ is bounded on the $SL(2,\mathbb{Z})$ fundamental domain and, by modular invariance, globally. We have
\begin{equation}
\cZ^{(0)}(\b)\coloneqq
\sum_{\lambda>0}a_\lambda e^{-\beta\lambda}, \qquad |\mathcal\cZ^{(0)}(\b)| \le{\sqrt{2\pi \o\beta}}\,\sup_{\tau\in\mathbb H}|\cZ(\tau)|<\infty
\end{equation}
for any $\b>0$, where the scalar partition function $\cZ^{(0)}(\b)$ is defined by the projection \eqr{Z0y}. We decompose $\cZ^{(0)}(\b)$ into edge, background, and fluctuation terms. The finite edge contributes $O(1)$ as $\beta\rar 0$, while, by Abel summation, $\delta\cZ^{(0)}_{\bulk}(\beta) =\beta\int_{\lb}^{\infty}\delta S(\l)e^{-\beta \l}\,d\l =O(\beta^{-a})$. Thus the reference transform satisfies
\begin{equation}
\begin{aligned}
\cZ^{(0)}_{\overline{\bulk}}\,\,(\beta)
&\coloneqq
\beta\int_{\lb}^{\infty}\bar S(\l)e^{-\beta \l}\,d\l
=\cZ^{(0)}(\beta)-\cZ^{(0)}_{\edge}(\beta)
  -\delta\cZ^{(0)}_{\bulk}(\beta)
=O(\beta^{-a}+\beta^{-1/2})
\end{aligned}
\end{equation}
as $\b\rar 0$. Note that $\int_{\lb}^{\infty}|\bar S(\l)|e^{-\beta \l}\,d\l$ converges for every $\beta>0$.\foot{Let $\cN(\l)$ and $\widetilde \cN(\l)$ count the bulk scalar primaries of the two spectra in the interval $[\lb,\l)$, so $S(\l) = \cN(\l) - \widetilde \cN(\l)$. Each of these counts is upper-bounded by a bounded multiple of the integrated Cardy formula. Since $\bar S(\l)=S(\l)-\delta S(\l)$ and hence $|\bar S(\l)| \leq \cN(\l) + \widetilde\cN(\l)+|\delta S(\l)|$, the fluctuation bound $|\delta S(\l)|=O(\l^a)$ implies $|\bar S(\l)| \leq \exp[O(\sqrt{\l})]$ as $\l\rar\i$, which implies the stated result.} 

Now suppose that $\bar S(\l)$ is {\it eventually} sign-definite, i.e. that $\bar S(\l\geq R_c)$ is sign-definite for an $R_c<\infty$ which may depend arbitrarily on $c$. Then
\begin{equation}
\begin{aligned}
\beta\int_{\lb}^{\infty}
|\bar S(\l)|e^{-\beta \l}\,d\l
&\le
|\cZ^{(0)}_{\overline{\bulk}}\,(\beta)|
+2\beta\int_{\lb}^{R_c}|\bar S(\l)|\,d\l=O(\beta^{-a}+\beta^{-1/2})
\end{aligned}
\end{equation}
as $\b\rar 0$. Tonelli's theorem consequently gives
\begin{equation}\label{tonelli}
\begin{aligned}
\int_{\lb}^{\infty}
|\bar S(\l)|\l^{-H-1/2}\,d\l
&=\frac{1}{\Gamma(H+\tfrac12)}
\int_0^\infty \beta^{H-1/2}
\left[
\int_{\lb}^{\infty}
|\bar S(\l)|e^{-\beta \l}\,d\l
\right]d\beta&<\infty
\end{aligned}
\end{equation}
for $H>\max\{1,a+\tfrac12\}$, which ensures convergence at $\beta=0$. This is the range of $H$ quoted in the main text. Finally, the Mellin representations of $A(s)$, $A_{\edge}(s)$, and $\delta A_{\bulk}(s)$ imply
\begin{equation}
\begin{aligned}
\bar A_{\bulk}(s)
&=
\frac{\lambda_1^{s-1/2}}{\Gamma(s-\tfrac12)}
\int_0^\infty
\beta^{s-3/2}\cZ^{(0)}_{\overline{\bulk}}\,(\beta)\,d\beta\\
&=(s-\tfrac12)\lambda_1^{s-1/2}
\int_{\lb}^{\infty}
\bar S(\l)\l^{-s-1/2}\,d\l.
\end{aligned}
\end{equation}
This establishes the Abel formula \eqr{abelbar}.

We can now bring things together with the previous subsection. Under the enigmatic/asymptotic split made there, $\bar S(\l)$ is frozen up to polynomial fluctuations for all $\l$ in the asymptotic region -- that is, $|\bar S(\l) - \bar S(\l_a)| \lesssim c^b\l^a$ for $\l\geq \l_a$ -- so \eqr{tonelli} holds. The remaining case to treat is therefore the possibility of super-polynomial fluctuations of $\bar S(\l)$ that persist out to {\it arbitrarily} large $\l$, not confined to the enigmatic region. In particular, the putative case of physical relevance is if $\rho_0(\l)$ and $\widetilde\rho_0(\l)$ were to have the same {\it leading}-order Cardy asymptotic, but one of them differs at some subleading but still exponential order.\foot{For example, if one spectrum were log-rigid around the non-perturbative $S$-modular vacuum kernel in \eqr{Cardydef}, but the other were only log-rigid around the leading-order Cardy exponential.} What the above derivation shows is that precisely this situation is forbidden. This excludes persistent asymptotic exponential offsets between the two references, supporting the separation from Part \hyperref[appb1]{B.1} into an enigmatic region of potentially large differences and an asymptotic region of only polynomial differences. Conversely, the above derivation implies that exceptions can only occur if $\rho_0(\l)-\widetilde\rho_0(\l)$ were to enjoy persistent super-polynomial oscillations {\it of both signs}; while this possibility must be borne in mind, it is not compatible with any reasonable reference densities which, unlike exact microscopic densities, admit smooth asymptotic expansions without short-range oscillations.

\end{appendix}
\end{spacing}

\bibliographystyle{JHEP}
\bibliography{NMB_refs_final_v2}

\end{document}